%% file: TMG.tex
\documentclass[a4paper,11pt]{article}
\pdfoutput=1 

\usepackage{jheppub} 
                     
\usepackage[mathscr]{euscript}
\usepackage[T1]{fontenc} 
\usepackage[textsize=footnotesize,textwidth=2.5cm]{todonotes}

\usepackage{relsize}

\definecolor{mikadoyellow}{rgb} {0.16, 0.254, 0.6}

\usepackage{amssymb,amsmath,latexsym,bm,amsfonts}
\usepackage{graphicx}
\usepackage{longtable}
\usepackage{color,xcolor}
\usepackage{indentfirst}
\usepackage{subfigure}
 \usepackage{comment}
 \usepackage{float}

\input{ShortCommand}

 \title{ \boldmath Anomalous Gravitation and its Positivity from Entanglement    }

\author[a]{Hongliang Jiang } 

\affiliation[a]{Albert Einstein Center for Fundamental Physics,    Institute for Theoretical Physics,  University of Bern,  
 			Sidlerstrasse 5, 3012 Bern, Switzerland}
 
\emailAdd{jiang@itp.unibe.ch}

\abstract{We explore the emergence of gravitation from entanglement in  holographic CFTs with gravitational anomalies. More specifically, the holographic correspondence between topologically massive gravity (TMG) with gravitational Chern-Simons term in the 3D bulk and its dual  CFT with unbalanced left and right moving central charges on the 2D boundary, is studied  from the quantum entanglement perspective. Using the first law of entanglement, we derive the holographic dictionary of the energy-momentum tensor in TMG, including the chiral case  with logarithmic   mode. Furthermore, we show  that the linearized equation of motion of TMG can also be obtained from entanglement using the Wald-Tachikawa covariant phase space formalism.   Finally, we identify a quasi-local gravitational energy in the entanglement wedge as the holographic dual  of relative entropy in gravitationally anomalous CFTs. The positivity and monotonicity  of relative entropy imply that such a gravitational energy should be positive definite and become  larger when increasing the size of the entanglement wedge. These constraints from quantum information  may be potentially used to discuss the UV inconsistent issues of TMG. 

}

\begin{document}
\maketitle
\flushbottom

  \section{Introduction}

 As the only untamed force in nature,  gravity is ubiquitous but puzzling. Although the classical gravity is well described in the framework of general relativity, its full-fledged quantum regime remains mysterious.  Holography opens a window to understand quantum gravity by associating it with  a boundary field theory which is better understood \cite{tHooft:1993dmi,Susskind:1994vu,Maldacena:1997re,Witten:1998qj}.  Especially, recent studies found a compelling relation between the geometry of bulk spacetime and the entanglement patter of the boundary field theory.  The RT-HRT proposal shows that the boundary entanglement entropy is given by the area of the minimal/extremal surface in the bulk spacetime \cite{Ryu:2006bv,Hubeny:2007xt}.  In spite, understanding the emergence of spacetime from entanglement is still   obscure. A key step towards this direction is made possible by deriving  explicitly the  equation of motion, which governs the classical gravity, from entanglement. More specifically, in the framework of AdS/CFT correspondence, the authors in    \cite{Lashkari:2013koa} derived the  linearized Einstein equation in AdS  from entanglement.   The generalizations to higher derivative gravities and     Einstein equation  at  the non-linear level were  also considered later in \cite{Faulkner:2013ica, Faulkner:2017tkh,Haehl:2017sot}. This provides a promising and direct way to help understand how the bulk gravity is assembled in the field theory. It is thus important to push the ideas further. The main goal of this paper is  to  generalize their discussions and try to understand the emergence of gravitation from    CFTs with  gravitational anomalies. 

More specifically, we consider  the topologically massive gravity (TMG) in   three dimensional AdS bulk. The Chern-Simons (CS) term  in TMG accounts for the unbalanced central charges of the left- and right- movers in the boundary 2D CFT,     thus rendering  the  gravitational anomalies on the boundary.  For such an AdS${_3}$/CFT$_{2}$ correspondence with gravitational anomalies, the holographic entanglement entropy (HEE) is given by the sum of the geodesic length   and the twist of the normal frame along the geodesic \cite{Castro:2014tta}.  Based on this proposal and using the first law of entanglement, we derive the holographic dictionary of stress-tensor in TMG taking into account the gravitational anomalies on the boundary. The dictionary agrees with the one derived using holographic renormalization procedure \cite{Skenderis:2009nt} which is quite complicated. Especially, the  holography dictionary of stress-tensor of TMG at the chiral point is also obtained automatically by combining  the   entropy relation and Lorentz invariance. Making further use of the  first law of entanglement, we also derive the linearized equation of motion in TMG purely from the entropic considerations. The bridge  relating the two sides is the Wald-Tachikawa covariant phase space formalism \cite{Wald:1993nt,Iyer:1994ys,Tachikawa:2006sz}.

We then go beyond the linear order first law of entanglement  and consider the relative entropy. We find the holographic  dual of the relative entropy in boundary CFTs with gravitational anomalies, which is given by a vacuum-substrated quasi-local energy  in the entanglement wedge. The discussions thus generalize   the pure Einstein gravity \cite{Lashkari:2016idm} to TMG. The relative entropy is known to be positive  definite and monotonically increasing with subregion size.  The holographic correspondence thus translates these quantum information inequalities into positive energy theorems in the bulk. By virtue of these positive energy theorems,  the quasi-local energy in the entanglement wedge is positive and increases when making the entanglement wedge larger.  Any   low energy effective field theory which  violates such  generalized positive energy theorems can not be UV completed in quantum gravity.  Therefore the quantum information theory provides some criteria  of fencing in the swampland.\footnote{The story can also be reversed. For example,  the 
monogamy of mutual information \cite{Hayden:2011ag} based on RT proposal is not   true in general, but  does hold for holographic states. This thus also offers a way to chart the space of holographic CFTs. In this paper, we always assume that the boundary  field theories are holographic, admitting large-$N$ limit and spare spectrum/a large gap.  }
It has been argued in \cite{Maloney:2009ck,Li:2008dq} that the TMG itself is unstable/inconsistent  generically due to the negative energy of either massive gravitons or BTZ black holes,  and  is  thus in the swampland generically.  The only possible UV completable TMG is chiral gravity with $\mu\ell=1$. A potential   application of these inviolable quantum information inequalities may thus be to show the UV inconsistence of non-chiral TMG. We leave it as  a future   direction.

This paper is organized as follows. In section \ref{HEETMG}, we review the topologically massive gravity, its holographic dual and the holographic entanglement entropy proposal. We will also derive the variation of the HEE for later usage.  In section~\ref{holoDictionary}, we derive  the holographic dictionary of stress tensor  purely from entanglement. In section~\ref{WTformalism}, we review the Wald-Tachikawa covariant phase space formalism. In section~\ref{eomTMG}, we derive the linearized equation of motion  in TMG based on entropic considerations. In section~\ref{relEnt}, we consider  the relative entropy and obtain its holographic dual. In section~\ref{conclu}, we conclude and discuss possible directions for further studies.  In appendix~\ref{appRindler},  
we give  the explicit expressions of the modular flow generator in general BTZ background. As a byproduct,  we  also derive  the HEE of TMG in Poincare AdS using Rindler method.  In appendix~\ref{EEphase},  we compute the entanglement entropy by integrating the charge in phase space. This general method is supposed to be applicable for calculating  the  entanglement entropy  in   general holographic  setups.

   \section{HEE in AdS${_3}$/CFT$_{2}$  with gravitational anomalies   }\label{HEETMG}
   
   \subsection{Topologically massive gravity }
The action of TMG  in AdS$_3$ is given by   the sum of Einstein-Hilbert term, cosmological constant term and the 
Chern-Simons term \footnote{
We use the following convention:  
\be
 dx^\lambda \wedge dx^\mu \wedge dx^\nu    
 =   \sqrt{-g}    \epsilon^{\lambda\mu\nu} d^3 x~,
\qquad
\epsilon_{z tx }=\sqrt{-g}~, \qquad \epsilon^{ztx } =-\frac{1}{\sqrt{-g}}~.
\ee
}
 \footnote{
 The minus sign in front of the CS term  comes from different convention of anti-symmetric tensor $\epsilon$. With this minus sign, the Lagrangian here is the same as the one in \cite{Skenderis:2009nt} in component form. 
 }
   \beqn
   S&=&\frac{1}{16\pi G_N} \int d^3 x \; \sqrt{-g} \Big(R-\Lambda\Big)
-\frac{1}{32\pi G_N \mu } \int  \tr\Big(  {\bf \Gamma} \wedge  d{\bf \Gamma}+\frac23 {\bf \Gamma}\wedge{\bf \Gamma}\wedge{\bf \Gamma} \Big)
\\&=&
\frac{1}{16\pi G_N} \int d^3 x \; \sqrt{-g} \Big(R+\frac{2}{\ell^2}\Big)
   -\frac{1}{32\pi G_N \mu } \int d^3 x \; \sqrt{-g}
    \epsilon^{\lambda\mu\nu} \Gamma^\rho_{\lambda\sigma} 
    \Big( \p_\mu \Gamma^\sigma_{ \nu\rho}+\frac23 \Gamma^\sigma_{\mu\tau} \Gamma^\tau_{\nu\rho} \Big)~,
    \qquad\quad
   \eeqn
  where we introduce the   matrix-valued connection  $ {   \bf \Gamma}^\mu_{\;\;\nu}=\Gamma^\mu_{\rho\nu} dx^\rho $. 
 
%
%
%

The equation of motion of TMG is 
\be\label{eomTensor}
\mathcal E_{\mu\nu}\equiv \frac{1}{16\pi G_N} \Big(
  R_{\mu\nu}-\frac12 g_{\mu\nu} R-\frac{1}{\ell^2} g_{\mu\nu} -  \frac{1}{\mu} C_{\mu\nu} \Big)=0~,
\ee
where the Cotton tensor $C_{\mu\nu} $ is defined as  
\be
C_{\mu\nu} = \epsilon_\mu^{\;\; \alpha\beta} \nabla_\alpha\Big( R_{\beta\nu}-\frac{1}{4} g_{\beta\nu}R\Big)
= \frac{1}{2} \Big(  \epsilon_\mu^{\;\; \alpha\beta}   \nabla_\alpha R_{\beta \nu}  + \epsilon_\nu^{\;\; \alpha\beta}   \nabla_\alpha R_{\beta \mu}   \Big)~,
\ee
and has the following properties
\be
\epsilon^{\alpha\mu\nu}C_{\mu\nu}=\nabla^\mu C_{\mu\nu}=C^\mu_{\;\; \mu}=0~.
\ee

As a consequence, all solutions of TMG have constant scalar curvature $R=-6/\ell^2$ and the equation of motion can also  be rewritten as
\be
R_{\mu\nu}+\frac{2}{\ell^2} g_{\mu\nu}= \frac{1}{\mu}C_{\mu\nu}~.
\ee 

It is consistent to set $C_{\mu\nu}=0$. \footnote{TMG also admits novel solutions with $C_{\mu\nu} \neq 0 $ which are not locally AdS$_3$, but we will not study them here. }
 Then the equation of motion of TMG essentially reduces to Einstein equation and thus its solution is locally AdS$_3$.  TMG can then be discussed in the context of  AdS$_3$/CFT$_2$ correspondence. 
    Specializing   to the locally AdS$_3$ solutions with Brown-Henneaux boundary conditions, the asymptotic symmetry analysis implies the dual CFT has two copies of Virasoro symmetry with central charges
        \be
    c_L=\frac{3\ell}{2G_N} \Big( 1+\frac{1}{\mu}\Big)~, \qquad     c_R=\frac{3\ell}{2G_N} \Big( 1-\frac{1}{\mu}\Big)~,
    \ee
or equivalently 
 \be
 \frac{\ell}{4G_N \mu}=\frac{c_L-c_R}{12}~, \qquad \frac{\ell}{4G_N}=\frac{c_L+c_R}{12}~.
 \ee
 
The  unbalanced  central charges indicate  the gravitational anomalies of the boundary field theory  and  thus renders the non-conservation of boundary stress-tensor. The gravitational anomalies  can also be understood from the bulk: under the diffeomorphism, the  gravitational Chern-Simons  term  in the bulk is only  invariant  up to a boundary term, which gives rise to the non-vanishing divergence of the stress tensor in  the boundary field theory. 
 
Before closing this subsection, we would like to emphasise the special case $\mu\ell=1$, the so-called ``chiral point''. In this case $c_R=0$ and the bulk gravity is chiral gravity which is holographically dual to chiral CFT \cite{Li:2008dq}. On the hand, the boundary conditions  at the chiral point  could be relaxed to admit  logarithmic   mode  \cite{Grumiller:2008qz}.  The resulting gravity is called log gravity and the dual boundary field theory is believed to be  logarithmic CFT \cite{Grumiller:2008qz,Maloney:2009ck}. In spite, the holography dual at the chiral point remains controversial.  As argued in \cite{Maloney:2009ck,Li:2008dq}, actually only the chiral gravity is possibly UV completable.  Other TMGs are inconsistent due to the negative energy of either massive gravitons or BTZ black holes. In this paper, we temporarily ignore the UV issues of TMG, but we will comment  on this point in the discussion of relative entropy. 
  
 In the remainder of the paper, we will set the AdS radius $\ell=1$.

\subsection{HEE in TMG}
\begin{figure}[h] 
   \centering
    \includegraphics[width=0.25\textwidth]{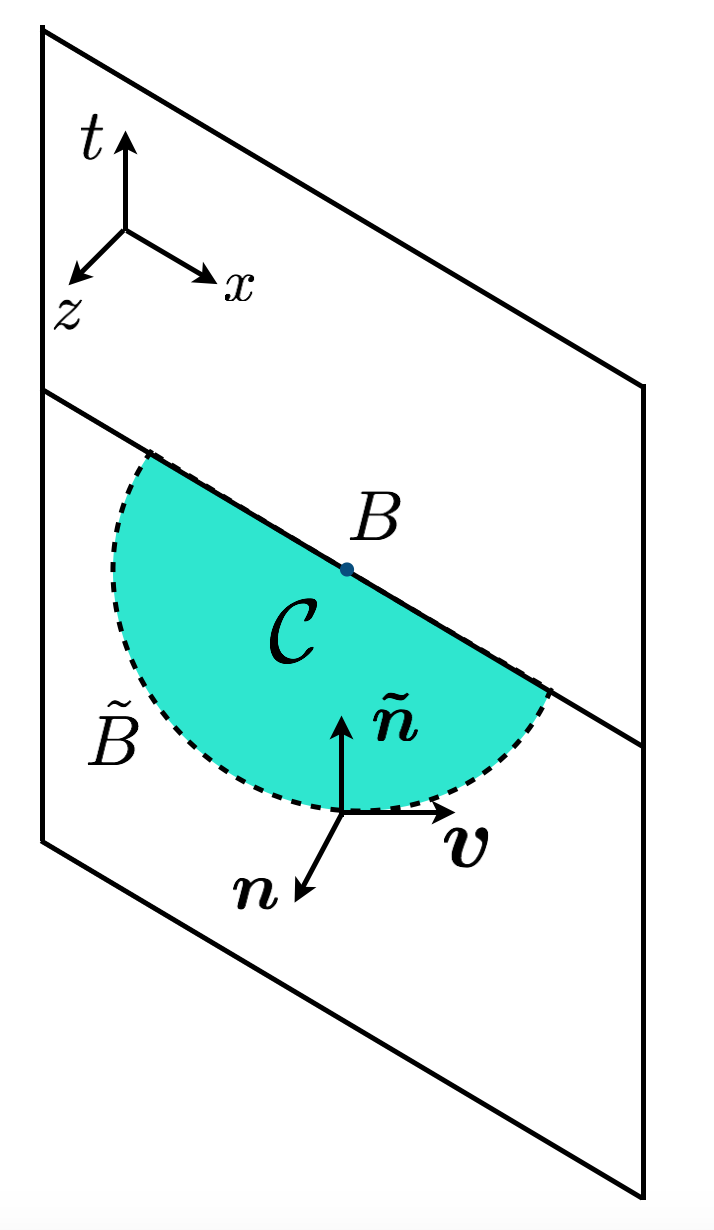} 
    \caption{
    The geometric picture of HEE in TMG. $B$: the boundary interval, $\tilde B$: the geodesic homologous to $B$, $\mathcal C$: the  surface enclosed by $B$ and $\tilde B$,   $(\bm v, \bm n, \bm {\tilde n})$: the normal frame along the geodesic. The plane is the boundary of AdS and CFT lives there.  
    }
\label{plot} 
\end{figure}

For pure Einstein gravity,  the famous RT-HRT proposal  \cite{Ryu:2006bv,Hubeny:2007xt} provides a holographic way to calculate the entanglement entropy and it  is given by the area of  the minimal or extremal surface in the bulk.  For  AdS$_3$, it is just the length of the bulk geodesic connecting the endpoints of the boundary interval. However, once including the Chern-Simons term in gravity, the extremization prescription is modified \cite{Castro:2014tta}:
\be
S_\text{HEE}=\text{ext}_\gamma\quad \frac{1}{4G_N}\int_{\gamma}  ds\Big(\sqrt{g_{\mu\nu} \dot X_\mu \dot X_\nu}
  + \frac{1}{\mu}g_{\mu\nu} \tilde n^\mu v^\rho \nabla_\rho n^\nu  \Big)~,
\ee
where $s$ is the proper length and we introduced the normal frame along $\gamma$ (see figure~\ref{plot})   
\be
v^\mu=\frac{dX^\mu}{ds},\quad
\tilde n^\mu =\epsilon^{\mu\nu\rho} v_\nu n_\rho,
 \quad \bm v\cdot \bm n=\bm v\cdot \tilde{ \bm n}=\bm n\cdot  \tilde {\bm   n}=0, \quad
\bm v\cdot\bm  v= \tilde{\bm  n}\cdot \tilde {\bm n}=-\bm n\cdot\bm  n=1 ~.
\ee

If the bulk spacetime is locally AdS$_3$, the curve $\gamma$ after doing extremization  is actually the geodesic. So the entanglement entropy of the boundary interval $B$ is given by the sum of   length  of the geodesic and twist of the normal frame along  the geodesic:
\beqn\label{HEEproposal}
S_\text{HEE}&=& \frac{ \text{\sf  Length} }{4G_N} +\frac{  \text{\sf  Twist} }{4G_N \mu} 
\\&=&
\frac{1}{4G_N}\int_{\tilde B}  ds \sqrt{g_{\mu\nu} \dot X_\mu \dot X_\nu}
  + \frac{1}{4G_N \mu }\int_{\tilde B}  ds\; g_{\mu\nu} \tilde n^\mu v^\rho \nabla_\rho n^\nu ~.
  \eeqn
%
%
%
%
%
%
%

As applications, we consider the EE in thermal CFTs which are holographically dual to BTZ black holes.   
The standard metric of  BTZ black hole takes the form  
\be\label{BTZ}
 ds^2=- \frac {(\rho^2-\rho_+^2)(\rho^2-\rho_-^2)}{\rho^2} d\tau^2+
  \frac{\rho^2}{(\rho^2-\rho_+^2)(\rho^2-\rho_-^2)} d\rho^2 
 +\rho^2 (d\chi +\frac{\rho_+\rho_-}{\rho^2}  d\tau )^2~,
 \ee
 where $\rho_\pm$ are the radial position of the outer and inner horizons and they are related to the mass and angular momentum of black holes.
 
 Consider the following interval  on  the boundary plane $( \tau, \chi)$
\be\label{intervalchitau}
B:\qquad (\tau=0,\chi=-R_\chi )\rightarrow (\tau=0,\chi= R_\chi )~.
\ee  
 where we used the right arrow symbol to indicate  an interval connecting the two endpoints. 
 
 Because of the factorization of left-moving and right-moving modes, the EE of this subregion is contributed by the sum of these two sectors. On the other hand, the EE can be calculated using the proposal we reviewed above. Indeed they agree and the result is \cite{Castro:2014tta}
 \beqn
  S_\text{EE}&=&S_{L}+S_{R}=
  \frac{c_L }{6}\log 
 \Bigg( \frac{\beta_L  }{\pi  \varepsilon_\chi }  \sinh\Big(\frac{2\pi R_\chi }{\beta_L}\Big)    \Bigg)
 +
\frac{c_R }{6}\log 
 \Bigg( \frac{\beta_R }{\pi  \varepsilon_\chi }  \sinh\Big(\frac{2\pi R_\chi }{\beta_R}\Big)    \Bigg)
\\ &=&
 S_\text{RT}+S_\text{CS}=\frac{c_L+c_R}{12}\log 
 \Bigg( \frac{\beta_L \beta_R}{\pi^2 \varepsilon_\chi^2}  \sinh\Big(\frac{2\pi R_\chi }{\beta_L}\Big)  \sinh\Big(\frac{2\pi R_\chi }{\beta_R}\Big)   \Bigg)
\\ & &\qquad\qquad\qquad+
 \frac{c_L-c_R}{12}\log 
 \Bigg(   \frac{\beta_L   \sinh\Big(\frac{2\pi R_\chi }{\beta_L}\Big)  }{\beta_R  \sinh\Big(\frac{2\pi R_\chi }{\beta_R}\Big)}   \Bigg)~,
 \eeqn
where the temperature of the left-mover and right-mover in CFT are related to the mass and angular momentum of BTZ black hole. More precisely, they are
 \be
 \beta_L=\frac{2\pi}{\rho_+-\rho_-}, \qquad  \beta_R=\frac{2\pi}{\rho_++\rho_-}~.
 \ee
 
 For our convenience, we would like to translate the results above into another coordinate system.   
 The phase space of AdS$_3$ solutions are given by 
 \be
 ds^2=\frac{dr^2}{4r^2}+\Big(2r+ \frac{U(u) V(v) }{2r} \Big) du dv+\Big( U(u) du^2+V(v)dv^2\Big)~.
 \ee
With this coordinate, the factorization of left- and right- moving sectors becomes manifest. Wick  rotating to the Euclidean signature, the light-cone coordinates  $u,v$ become the complex coordinate $z, \bar z$. 
 
We are especially interested in the  zero mode   cases  which correspond  to constant functions $U,V$ and admit  high symmetries 
   \be\label{zeroGeometry}
 ds^2=\frac{dr^2}{4r^2}+\Big(2r+ \frac{U  V  }{2r} \Big) du dv+\Big( U  du^2+V dv^2\Big)~.
 \ee
 They are essentially the BTZ black holes \eqref{BTZ}.  The two coordinate systems are related through
 \be\label{tsfScale}
 \rho=\frac{\sqrt{(r+\rho_-^2)^2+2(r-\rho_-^2)\rho_+^2+\rho_+^4}}{2\sqrt{r}}, 
 \quad  \chi=\sqrt{2}(u+v), \quad \tau=\sqrt{2}(u-v)~,
 \ee
and 
 \be
 U=2(\rho_++\rho_-)^2, \qquad V=2(\rho_+-\rho_-)^2~. 
 \ee
 
The horizons of the black holes in our coordinates sit at 
\be
r_+=\rho_+^2-\rho_-^2=\frac{1}{2} \sqrt{UV}, \qquad r_-=-\rho_+^2+\rho_-^2=-\frac{1}{2} \sqrt{UV}~.
\ee

 We would like to express the  results of EE in terms our coordinate systems. The boundary plane $(u,v)$ and $(  \tau,\chi)$  are related through a nontrivial rescaling \eqref{tsfScale}.  The interval in \eqref{intervalchitau} in our new coordinate system $( \tau,\chi)$ is expressed as
    \be\label{intervalconstT}
B:\qquad (u=-R, v=-R) \rightarrow (u= R, v=R)~.
   \ee
Then it is easy to see
\be
 2 R_\chi = l_\chi=\sqrt{2} (l_u+l_v)=4\sqrt{2} R ~, \qquad l_\tau=\sqrt{2} (l_u+l_v)=0~,
  \ee
yielding 
 \be
 R_\chi = \sqrt{8} R~.
 \ee

With these relations, we can easily translate the results of EE into our new coordinate system 
  \beqn\label{EEnewcoord}
 S_\text{EE}=S_\text{RT}+S_\text{CS}&=&\frac{c_L+c_R}{12}\log 
  \Bigg( \frac{8}{\sqrt{UV}\varepsilon_\chi^2}   \sinh\Big( 2R \sqrt{U}   \Big)    \sinh\Big( 2R \sqrt{V}   \Big)     \Bigg)
\\ &+&
 \frac{c_L-c_R}{12}\log 
  \Bigg(   \frac{\sqrt{V}  \sinh\Big( 2R \sqrt{U}   \Big)  }{\sqrt{U}  \sinh\Big( 2R \sqrt{U}  \Big)}   \Bigg)~.
  \label{EECSnew}
 \eeqn
Consider the limit that $U,V$ are small perturbations, then 
 \be\label{SEEpert}
 S_\text{EE}=\frac{1}{2G} \Big( \log \frac{R}{\varepsilon} +\frac{R^2}{3} (U+V) \Big) 
 +\frac{1}{6G\mu}R^2(U-V)+ o(U,V)~,
 \ee
 where $\varepsilon=\varepsilon_\chi/\sqrt{32}$ and we used the dictionary of central charges. 
 
As a byproduct of this paper, in the appendix~\ref{EEphase} we will derive the entanglement entropy using other two different approaches: Rindler method and integration in the phase space. The second method applies to arbitrary zero mode background (namely arbitrary temperature CFT) and arbitrary interval (not necessarily on the constant time slice). These results reduce to the known results in the proper limit. The key point is regarding the entanglement entropy, like the black hole entropy, as a Noether charge. This is quite generic and should be applicable to much  wider classes, especially to those non-AdS holography.

  \subsection{Symmetries and modular flow}

For later discussions, we also need to understand the symmetries and the modular flow.    
   
Consider the following   interval on the   boundary of AdS (in  Minkowski coordinate $(t,x)$)
   \be\label{intervalconstT}
B:\qquad (  0,  -R) \rightarrow (0, R)~,
   \ee
 
For every interval, one can associate a  Killing vector generating the modular flow. This modular flow  generator can be obtained through the Rindler method. Especially for the interval in  \eqref{intervalconstT}, it is \cite{Casini:2011kv}
 \be\label{modflow1}
 \xi_B=-\frac{2\pi t}{R} (z\p_z+x \p_x)+\frac{\pi}{R} (R^2-z^2-t^2-x^2) \p_t~.
 \ee   
 Restricting to the boundary of AdS, it reduces to 
 \be\label{boundaryflow}
\zeta_B=-\frac{2\pi t x}{R}     \p_x +\frac{\pi}{R} (R^2 -t^2-x^2) \p_t~.
 \ee 
   
More generally, one can   study the boosted interval. It is convenient to introduce the light-cone coordinates due to the factorization of the left and righer movers
 \be\label{vacuumGeometry}
 ds^2=G^{(0)}_{\mu\nu}dx^\mu dx^\nu=\frac{dz^2 -dt^2+dx^2}{z^2} =\frac{dr^2}{4r^2}+2r du duv~,
 \ee  
The two coordinate systems are related through
  \be
 r=\frac{1}{2z^2}, \qquad u=x+t, \qquad v=x-t~.
 \ee

 The most general subregion  of interest is then given by
   \be
 - (\frac{l_u}{2},\frac{l_v}{2}) \rightarrow (\frac{l_u}{2},\frac{l_v}{2})~. 
   \ee
The associated Killing vector can also be obtained through Rindler method. See appendix~\ref{appRindler} for details where we also work out the modular flow in the most general BTZ background. The  result is 
\beqn 
\xi&=&-  \frac{2\pi}{l_u} L_1 + \frac{\pi l_u}{2}  L_{-1} + \frac{2\pi}{l_v}  \bar L_1 -   \frac{\pi l_v}{2} \bar L_{-1} \\
&=& 4 \pi r \Big(\frac{u}{l_u}-\frac{v}{l_v}\Big)\p_r +\frac{\pi}{2} \Big( l_u -\frac{2}{r l_v }-\frac{4u^2}{l_u} \Big) \p_u
-\frac{\pi}{2} \Big( l_v -\frac{2}{r l_u }-\frac{4v^2}{l_v} \Big) \p_v~.
\label{modflow2}
\eeqn
On the boundary, it reduces to 
\be
\zeta=\frac{\pi}{2} \Big( l_u -\frac{4u^2}{l_u} \Big) \p_u
-\frac{\pi}{2} \Big( l_v  -\frac{4v^2}{l_v} \Big) \p_v  ~.
\ee

It is easy to see that  at the endpoints of the interval $r=\infty, u=\pm \frac{l_u}{2}, v=\pm \frac{l_v}{2}$, the modular flow vanishes $\zeta=\xi=0$. 
The unboosted interval corresponds to $l_u=l_v=2R$ and the above general form reduces to the special case. 
 
The RT-HRT minimal surface, more precisely the geodesic here, coincides with the fixed points of the modular flow  $\xi=0$:
 \be
 u=l_u \sqrt{\frac14-\frac{  1}{2r l_u l_v} } ~, \qquad   v=l_v \sqrt{\frac14-\frac{  1}{2r l_u l_v} }~.
 \ee
 In the special case of constant time slice $l_t=0$,
 \be
 l_u=l_v=2R:\qquad x^2+z^2=R^2 ~.
 \ee
 
%

\subsection{Variation of entanglement entropy }

In this subsection, we want to investigate the variation of entanglement entropy around the vacuum state whose dual geometry is the Poincare AdS \eqref{vacuumGeometry}. The variations of the vacuum state  correspond to the fluctuations of the bulk metric. 

First we consider the    Einstein-Hilbert  contribution  to the variation of EE. Using the RT-HRT proposal for HEE, we should consider the variation of geodesic length in the bulk. Since the  geodesic  is an extremal curve in the bulk,  to  leading order,  the position  of geodesic should not be modified and the only contribution comes from the change of induced metric along the geodesic \cite{Lashkari:2013koa}. So to leading order, the variation of EE from Einstein-Hilbert  term is 
  \be
 \delta S_\text{RT} = \delta  \frac{1}{4G_N}\int_{\tilde B} ds= \delta \frac{1}{4G_N} \int_{\tilde B} d\lambda \sqrt{G_{\mu\nu}  \dot  x^\mu \dot x^\nu}
 = \frac{1}{8G_N} \int_{\tilde B}  d\lambda  \frac{  \delta G_{\mu\nu}  \dot  x^\mu \dot x^\nu  }{ \sqrt{G^{(0)}_{\mu\nu}  \dot  x^\mu \dot x^\nu}}~,
 \ee 
 where $\tilde B$ is the geodesic in the unperturbative background.

For the CS term, we begin with a condition in the normal frame  $n\cdot \tilde n =n^\nu \tilde n_\nu =0$, which immediately implies that 
 \be
 \nabla_\rho(n^\nu \tilde n_\nu )=n^\nu  \nabla_\rho\tilde n_\nu +\nabla_\rho n^\nu \tilde n_\nu =0~,
 \ee
This enables us to rewrite 
 \be
 S_\text{CS}=\frac{1}{4G_N \mu } \int_{\tilde B} ds \; v^\rho \tilde n_\nu \nabla_\rho n^\nu
 =\frac{1}{8G_N \mu } \int_{\tilde B} ds \; v^\rho \Big(  \tilde n_\nu \nabla_\rho n^\nu -  n_\nu \nabla_\rho  \tilde n^\nu   \Big)~.
 \ee

 Now consider the fluctuation of the background metric. As in  Einstein-Hilbert case,  to leading order, the position of the geodesic is not modified and  the only  variation of EE in CS term is accounted for by the variation of the induced metric along the geodesic. More precisely we only need to consider the variation of    covariant derivatives $\delta \nabla=\delta(\partial+\Gamma)=\delta \Gamma$. 
 
Therefore
 \beqn
\delta S_\text{CS}& =&\frac{1}{8G_N \mu } \int_{\tilde B} ds \; v^\rho \Big(  \tilde n_\nu \delta\Gamma_{\rho\sigma}^\nu n^\sigma -   n_\nu \delta\Gamma_{\rho\sigma}^\nu \tilde n^\sigma    \Big)
\\& =&
\frac{1}{8G_N \mu } \int_{\tilde B} ds \; v^\rho\delta\Gamma_{\rho\sigma}^\nu g_{\mu\nu} \Big(  \tilde n^\mu   n^\sigma -   n^\mu    \tilde n^\sigma    \Big)~.
 \eeqn
 
Since $\tilde B$ is the bifurcate horizon of modular flow generator  $\xi_B$, we have  
 \be\label{binormal}
\text{on } \tilde B:\qquad \xi_B=0, \qquad \epsilon^{\mu \sigma} \equiv\tilde n^\mu n^\sigma -  n^\mu  \tilde n^\sigma  
=\frac{1}{\kappa_{\xi_B}}\nabla^\mu \xi_B^\sigma ~, \qquad
 \ee
where $\epsilon^{\mu \sigma}$ is called binormal and the surface gravity on the Killing horizon is defined as
 \be
\xi^\alpha \nabla_\alpha \xi_\beta =\kappa  \xi_\beta~.
 \ee
 Thus    to leading order, the variation of EE from  CS  term is 
 \beqn\label{deltaScs}
\delta S_\text{CS}  & =&\frac{1}{8G_N \mu } \int_{\tilde B} ds \; v^\rho\delta\Gamma_{\rho\sigma}^\nu g_{\mu\nu} \epsilon^{\mu\sigma}
=\frac{1}{8\pi G_N \mu\kappa_{\xi_B} } \int_{\tilde B} ds \; v^\rho\delta\Gamma_{\rho\sigma}^\nu   \nabla_\nu\xi_B^\sigma
\\&=&
\frac{1}{8\pi G_N \mu\kappa_{\xi_B} } \int_{\tilde B} dx^\rho\delta\Gamma_{\rho\sigma}^\nu   \partial_\nu\xi_B^\sigma~.
 \eeqn

For the modular flow generator in  \eqref{modflow1} and \eqref{modflow2}, the surface gravity is $\kappa=2\pi$.   So
\beqn\label{deltaScs}
\delta S_\text{CS}  & =&\frac{1}{8G_N \mu } \int_{\tilde B} ds \; v^\rho\delta\Gamma_{\rho\sigma}^\nu g_{\mu\nu} \epsilon^{\mu\sigma}
=\frac{1}{16\pi G_N \mu } \int_{\tilde B} ds \; v^\rho\delta\Gamma_{\rho\sigma}^\nu   \partial_\nu\xi^\sigma ~.
 \eeqn
 This expression coincides with the infinitesimal version of black hole entropy  from CS term \cite{Tachikawa:2006sz}.  Note that this is a covariant expression. Although the Christoffel symbol is not a tensor, its variation indeed is a tensor.      This expression can also be derived using the Rinlder method which maps the entanglement entropy to thermal entropy and  geodesic (a bifurcate horzion of modular flow)  to the  black hole  horizon. The covariant properties of the  expression guarantee that the form is the same before and after Rindler transformation. 
 
As a check of our proposal, we can consider the zero mode fluctuations around the vacuum. Then the fluctuated metric is given by \eqref{zeroGeometry} with $U, V$ being small. Using our formulae above, we easily find that
 \beqn\label{CScharge}
\delta S_\text{CS} &=& \frac{1}{4G \mu} \frac{1}{4\pi} \int_{\tilde B} ds\;     v^\rho \delta\Gamma^\nu_{\sigma\rho}  \p_\nu \xi^\sigma
 = \frac{1}{6G\mu}R^2(U-V) ~,
\\
 \delta S_\text{RT}
  &=&\frac{1}{4G} \int_{\tilde B} dr  \; \delta \sqrt{\frac{1}{4r^2}+\Big(2r+ \frac{U  V }{2r}+U+V \Big)( \frac{du}{dr})^2}
  =  \frac{R^2}{6G} (U+V) ~.
  \label{RTcharge}
  \eeqn
 
We   see that our perturbative calculations here agree with the exact result \eqref{SEEpert}.  In the following sections, we will consider generic fluctuations which are not necessarily constant modes and explore the physical implications.

 \section{Holographic dictionary of stress tensor from entanglement}\label{holoDictionary}
 
 As an important ingredient in holography, the holography dictionary establishes a map between the boundary field theory and bulk gravity. The standard approach to build the holographic dictionary is using holographic renormalization.  Especially,  the energy-momentum tensor of boundary CFT is given by the metric fluctuation in the bulk AdS gravity. In the presence of the CS term/gravitational anomalies, the dictionary becomes more complicated and  the complete derivation was done in \cite{Skenderis:2009nt}. 
 
  In \cite{Faulkner:2013ica}, using holographic entanglement the authors found an alternative simple way to derive  the  dictionary of  energy-momentum tenor  in the  Einstein gravity and higher derivative gravity. In this section, we will generalize their method and derive the  dictionary in the presence of CS term/gravitational anomalies. For simplicity, we focus on 3D gravity. But the method here  could be easily generalized to higher dimensions.

\subsection{Holographic dictionary from the first law of entanglement }

  Consider the Fefferman-Graham expansion of asymptotic AdS$_3$ around the vacuum state
\be\label{metricFluctuation}
 ds^2=\frac{dz^2     -dt^2+dx^2}{z^2}  +h_{ij}(z,t,x)dx^i dx^j~.
\ee
In this paper, we restrict ourselves to the asymptotic locally AdS case, so  $h_{ij}(z,t,x) \rightarrow o(\frac{ 1}{ z^2})$ near the boundary. The dictionary with more general boundary conditions is considered in \cite{Skenderis:2009nt},  but we are not going to explore it here partly because the entanglement entropy beyond AdS is not well understood yet. 

We consider the following interval 
\be
B:\qquad (t=0,x=-R) \rightarrow (t=0,x=R)~.
\ee
 
The key point in the derivation is the first law of entanglement
\be
\delta \EV{H_\text{mod }} =\delta S_\text{HEE}~,
\ee
 which relates the variation of vacuum expectation value of modular Hamiltonian and the variation  of holographic entanglement entropy. Let us first consider the variation of  holographic entanglement entropy about the vacuum state. It has two contribution. The Einstein-Hilbert  contribution can be considered using the RT-HRT proposal, given by the variation of geodesic length. To leading order, as we discussed in the last section, it is
  \be
 \delta S_\text{RT} = \delta  \frac{1}{4G_N}\int_{\tilde B} ds
 = \frac{1}{8G_N} \int_{\tilde B}  d\lambda  \frac{  \delta G_{\mu\nu}  \dot  x^\mu \dot x^\nu  }{ \sqrt{G^{(0)}_{\mu\nu}  \dot  x^\mu \dot x^\nu}}~,
 \ee 
 where $\tilde B$ is the geodesic. 
While for  the CS term, as we derived in the last section, the contribution is given by 
\be 
 \delta S_\text{CS}= \frac{1}{4G \mu} \frac{1}{4\pi} \int_{\tilde B} ds\;     v^\rho \delta\Gamma^\nu_{\sigma\rho}  \p_\nu \xi^\sigma~.
\ee

  
For the metric fluctuation given in \eqref{metricFluctuation}, the explicit results are
   \beqn
   \delta S_\text{RT}&=&\frac{1}{4G} \int_{-R}^R dx \frac{ R^2-x^2}{2R} h_{xx}(z,0,x)~, \qquad z=\sqrt{R^2-x^2}~,
\\
    \delta S_\text{CS}&=& \frac{1}{4G \mu}  \int_{-R}^R dx \frac{ R^2-x^2}{2R}
    \Bigg(
     h_{tx}(z,0,x)+ z \partial_z h_{tx} (z,0,x)+ x\Big( \partial_x h_{tx}(z,0,x) -\partial_t h_{xx}(z,t,x)|_{t=0} \Big) 
    \Bigg) ~.     \qquad
   \eeqn
 And their sum is
   \beqn
   \delta S&=& \frac{1}{4G }  \int_{-R}^R dx \frac{ R^2-x^2}{2R}
    \Bigg( h_{xx}(z,0,x)+
   \frac{1}{\mu}  h_{tx}(z,0,x)
  \nonumber  \\&&\qquad\qquad\qquad 
   + \frac{1}{\mu}z \partial_z h_{tx} (z,0,x) 
  + \frac{1}{\mu}x\Big( \partial_x h_{tx}(z,0,x) -\partial_t h_{xx}(z,t,x)|_{t=0} \Big)
    \Bigg)~.
   \eeqn
  
We want to consider the limit $R\rightarrow 0$. In this case,  $x,z\sim R\rightarrow 0$ and
   \beqn
  \delta S   &  \xrightarrow { R\rightarrow 0} & \frac{1}{4G }  \int_{-R}^R dx \frac{ R^2-x^2}{2R} 
    \Bigg( h_{xx}(z,0,x)+
    \frac{1}{\mu} h_{tx}(z,0,x) 
\nonumber    \\&&\qquad\qquad\qquad 
   +  \frac{1}{\mu}z \partial_z h_{tx} (z,0,x)+\frac{1}{\mu} x\Big( \partial_x h_{tx}(z,0,x) -\partial_t h_{xx}(z,t,x)|_{t=0} \Big)
    \Bigg) \Bigg|_{z=x=t=0}
 \nonumber   \\&=& 
     \frac{1}{4G }  \int_{-R}^R dx \frac{ R^2-x^2}{2R}\times 
    \Bigg( h_{xx}(0,0,0)+
 \frac{1}{\mu}    h_{tx}(0,0,0)+ \frac{1}{\mu} z \partial_z h_{tx} (z,0,0)|_{z=0} \Big)
\nonumber       \\&=&
     \frac{R^2}{6G }  
         \Bigg( h_{xx}(0,0,0)+
   \frac{1}{\mu}  h_{tx}(0,0,0)+\frac{1}{\mu}  \frac{\partial}{\partial \log z} h_{tx} (z,0,0)|_{z=0} \Bigg)~.
   \eeqn
   At first glance, the last term in the bracket should be discarded because of  the divergence in  $\log z$   near the boundary. However,  we don't know the $z$-dependence of the function. In order to have a finite answer, there could be a $\log z $ linear dependence in  $h_{tx}$ to regulate  the divergence in $z$. So we can have the following general ansatz   
\be\label{metricepxand}
h_{ij}(z,x^k)=2\log z \,B_{ij} (x^k) +H_{ij} (x^k)+...~,
\ee
where the ellipsis denotes the higher order terms in $z$ which have no contribution in $\delta S$. 

Plugging  the ansatz back, we get 
\be
  \delta S      \xrightarrow { R\rightarrow 0}   
   \frac{R^2}{6G }  
         \Big( H_{xx} +   \frac{1}{\mu} H_{tx} +   2\frac{1}{\mu}  B_{tx}  +2\frac{1}{\mu}\log z(B_{xx} + \frac{1}{\mu}  B_{tx} )  \Big) \Big|_{z=t=x=0} ~.
\ee
In order to have a finite answer on the boundary, the coefficient of $\log z$ should vanish 
\be\label{Bxxtmu}
B_{xx}=-\frac{1}{\mu}B_{tx}~.
\ee
 In this case, we have
 \be
  \delta S      \xrightarrow { R\rightarrow 0}   
   \frac{R^2}{6G }  
         \Big( H_{xx}(x=t=0) + \frac{1}{\mu} H_{tx}(x=t=0) -   2 B_{xx} (x=t=0)   \Big) ~.
\ee

On the other hand,  the modular Hamiltonian in CFT  is given by 
 \be\label{modH}
 H=\int_B \zeta_B^\mu T_{\mu\nu}\epsilon^\nu=2\pi   \int_{-R}^R dx \frac{ (R^2-x^2) }{2R}T_{tt}~.
 \ee
And its infinitesimal variation is 
\be
\delta \EV{H} \xrightarrow { R\rightarrow 0}    2\pi   \int_{-R}^R dx \frac{ (R^2-x^2)T_{tt}}{2R}=\frac{4\pi R^2}{3} \delta \EV{T_{tt}(x=t=0)}~.
\ee 

 Using the first law of entanglement $\delta S=\delta \EV{H}$, we thus find 
 \be
  \EV{T_{tt}}= \frac{1}{8\pi G_N}  \Big(   H_{xx}+ \frac{1}{\mu}H_{tx} -2B_{xx}\Big)~.
 \ee
 
 We would like to generalize it to arbitrary Lorentz frame. This can be done by considering the boosted interval and repeat all the discussions again. Another way is to use the Lorentz invariance and covariantize the above formula. Note that $H_{xx}=H^x{}_x =H_{tt}-\eta_{tt}H^k {}_k$ and $H_{tx}=-\epsilon_t{}^x H_{xt}=-\epsilon_t{}^k H_{k t}$
 \footnote{The convention is  $\epsilon_t{}^x=\epsilon_x{}^t=\epsilon_ {tx}=-1$.}.
So
  \be
  \EV{T_{ij}}= \frac{1}{8\pi G_N}  \Big(   H_{ij}-\eta_{ij} H^k {}_k -2B_{ij} +2 \eta_{ij} B^k {}_k -    
   \frac{1}{2\mu}  \epsilon_i{}^k H_{kj} -  \frac{1}{2\mu}   \epsilon_j{}^k H_{ki}    \Big)~.
 \ee
 The tracelessness of   the energy-momentum tensor in CFT $T^i {}_i=0$ implies the traceless of metric fluctuation
 \be\label{traceless}
H^k {}_k =B^k {}_k =0~.
 \ee
This enables us to rewrite 
  \be
    \EV{T_{ij}}= \frac{1}{8\pi G_N}  \Bigg(   H_{ij}  -2B_{ij}   - 
\frac{1}{2\mu} \Big( \epsilon_i{}^k H_{kj}  +i \leftrightarrow j \Big)   \Bigg)~.
 \ee  
 However, this is not the end of story. 
 
 \subsection{Further constraints from Lorentz invariance}
 
From \eqref{Bxxtmu} and \eqref{traceless} we have
 \be\label{Brelation}
B_{xx}= B_{tt}=- \frac{1}{\mu} B_{xt}=-\frac{1}{\mu} B_{tx}~.
 \ee
This is actually inconsistent with the Lorentz invariance unless $\mu=\pm 1$. To see this, define the map:
\be
 \widetilde {}:S_{ij} \mapsto  \widetilde S_{ik}=\epsilon_i{}^k S_{kj}~.
\ee 
Especially this is a linear map, implying 
\be
 \widetilde B_{xx}=-B_{tx}=\mu B_{xx}~.
\ee
and furthermore
\be
 \widetilde{ \widetilde B}_{xx}=\mu  \widetilde B_{xx}=\mu^2 B_{xx}~.
\ee
On the other,  this map is an involution  $\; \widetilde{}{}\; ^2=1$, implying that $ \widetilde{  \widetilde B}_{xx} =B_{xx}$. So these two are inconsistent unless $\mu=\pm 1$.  These two values are equivalent up to an orientation. In this paper, without loss of generality, as usual  we just consider $\mu=1$.  
So in general, if $\mu\neq 1$, the logarithmic mode in the asymptotic expansion should vanish $B_{ij}=0$. The logarithmic mode at the chiral point $\mu=1$ is exactly the one discovered in  \cite{Grumiller:2008qz,Skenderis:2009nt}.

 In this case, the \eqref{Brelation} can be compactly written as
 \be
 \epsilon_i{}^k B_{kj}=B_{ij}~.
 \ee
 
  The conservation of   the stress tensor
  \footnote{Since we are considering  that the boundary is flat Minkowski with Lorentz coordinate system and thus vanishing Levi-Civita connection, the gravitational anomalies do not spoil the usual conservation laws of energy-momentum tensor.   }
   in CFT $\partial^iT _{ij}=0$ implies that \footnote{ The identity 
 \be
 H_{ij}=H_{pq}\epsilon_i{}^p \epsilon_j{}^q~,
 \ee
which states that the tensor is symmetric and traceless,  is useful to show this. }
 
 \be\label{Tconserved}
 \partial^i B_{ij}=0, \qquad  (\delta_i^k - \frac{1}{\mu} \epsilon_i{}^k) \partial^j H_{jk}=0~.
 \ee
 
 We can expand the first conservation explicitly 
 \beqn
0 &=& \partial^x B_{xx}+ \partial^t B_{tx}= \partial^x B_{xx}-\mu \partial^t B_{xx} ~,
\\
0&=&   \partial^x B_{xt}+ \partial^t B_{tt}= -\mu \partial^x B_{xx}+  \partial^t B_{xx}=-{\mu} \partial^x B_{xx}+ \frac{1}{\mu }\partial^x B_{xx}   ~.
 \eeqn
 So these equations are not consistent unless $\mu=\pm 1$. (This argument is defective since it allows another possibility $\partial^x B_{xx}=0 $ for general $\mu$).
 
 The second equation  in \eqref{Tconserved}  can be multiplied  by $\epsilon_l{}^i$ which yields 
 \be
 (\epsilon_l{}^i\delta_i^k - \frac{1}{\mu} \epsilon_l{}^i\epsilon_i{}^k) \partial^j H_{jk}
 =\epsilon_l{}^i \partial^j H_{ji} - \frac{1}{\mu} \delta_i{}^k  \partial^j H_{jk}
 = (\mu -\frac{1}{\mu}) \partial^j H_{jl}=0~,
 \ee
 where we use \eqref{Tconserved} repeatedly as well as the property of $\epsilon$. 
 
 So, we have 
 \be
 \mu=\pm 1: \qquad  (\delta_i^k \mp \epsilon_i{}^k) \partial^j H_{jk}=0~,
 \ee
 or 
 \be
\mu\neq 1:\qquad   \partial^j H_{j  k}=0~.
 \ee

 To summarize, based on  the first law of entanglement,  Lorentz invariance and requiring a finite energy-momentum tensor, we find that   
 \begin{itemize}
 \item $\mu=1$: the asymptotic  expansion is given by 
 \be
h_{ij}(z,x^k)=2\log z B_{ij} (x^k) +H_{ij} (x^k)+...
\ee
where $B$ should satisfy the condition 
 \be
 \epsilon_i{}^k B_{kj}=B_{ij}~.
 \ee
 This is exactly the logarithmic mode for TMG at the chiral point, discovered  in \cite{Grumiller:2008qz,Skenderis:2009nt}.
 The holographic dictionary between stress tensor and asymptotic metric is given by
  \be\label{hol1}
  \EV{T_{ij}}= \frac{1}{8\pi G_N}  \Bigg(   H_{ij}  -2B_{ij}   - 
\frac{1}{2 } \Big( \epsilon_i{}^k H_{kj}  +i \leftrightarrow j \Big)   \Bigg)~.
 \ee 
 
 The conservation of stress tensor enforces  
   \be
 \partial^i B_{ij}=0, \qquad  (\delta_i^k -\epsilon_i{}^k) \partial^j H_{jk}=0~.
 \ee
 which arises from  the EoM of   TMG as shown in \cite{Skenderis:2009nt}.

 \item $\mu\neq 1$: the asymptotic  expansion is given by 
  \be
h_{ij}(z,x^k)= H_{ij} (x^k)+...
\ee

The holographic dictionary   is given by
  \be\label{hol2}
  \EV{T_{ij}}= \frac{1}{8\pi G_N}  \Bigg(   H_{ij}  - 
\frac{1}{2 \mu} \Big( \epsilon_i{}^k H_{kj}  +i \leftrightarrow j \Big)   \Bigg)~.
 \ee 
 
  The conservation of stress tensor enforces 
    \be
  \partial^j H_{jk}=0~.
 \ee

 \end{itemize}

 \section{Wald-Tachikawa covariant phase space formalism}\label{WTformalism}

In this section, we review the covariant phase space formalism \cite{Wald:1993nt,Iyer:1994ys,Tachikawa:2006sz} which is  useful for later discussions on equation of motion and relative entropy.

In the presence of CS term, the theory is not diffeormorphic invariant anymore. Acting with  a vector field, besides the normal Lie derivative action,   there are also anomalous contributions on the boundary.  More specifically, $\delta_\xi$ and $\mathcal L_\xi $ are not the same anymore.   To account for the difference, we need to add the  anomalous contribution by hand. 

\subsection{General formalism}
 
We consider the bulk gravity is $(d+1)$-dimensional. The volume form is given by
\footnote{For our problem, $d=2$ and $   \epsilon_{ztx}=\sqrt{-g}$.}
\be
{\bm \epsilon }=\frac{1}{(d+1)!} \epsilon_{\mu_1 .. \mu_{d+1}} dx^{\mu_1}\wedge \cdots \wedge dx^{\mu_{d+1}}~.
\ee
We also define the following tensors
\be
{\bm \epsilon_\mu }=\frac{1}{d !} \epsilon_{\mu \mu_2 .. \mu_{d+1}} dx^{\mu_2}\wedge \cdots \wedge dx^{\mu_{d+1}}, \qquad
{\bm \epsilon_{\mu\nu}}=\frac{1}{(d-1) !} \epsilon_{\mu\nu \mu_3 .. \mu_{d+1}} dx^{\mu_3}\wedge \cdots \wedge dx^{\mu_{d+1}}, \quad \cdots~.
\ee

The  action   is given by  Lagrangian density  integrated over spacetime 
\be
S=\int \mathcal L \bm\epsilon=\int \bm L ~.
\ee
Its variation gives rise to 
\be
\delta \bm L =\bm E_\phi \delta \phi+ d\bm\Theta[\delta \phi]~,
\ee
where $\phi$ collectively denotes all the fields  including the metric and $\bm E_\phi$ is the equation of motion.  

In the presence  of the CS term,  the variation under a diffeomorphism generated by an arbitrary vector field $\xi$ acquires extra contributions, in addition to  the Lie derivative action:
\beqn
\delta_\xi \bm  L &=&  \mathcal   L_\xi \bm L +d \bm \Xi_\xi~,
\\
\delta_\xi \bm  \Theta &=& \mathcal L_\xi \bm  \Theta +d \bm \Pi_\xi~.
\eeqn

One can show that the two anomalous terms are related
\be
d \bm\Pi_\xi = \delta d\bm \Xi_\xi~,
\ee
which implies that 
\be\label{PXS}
\bm \Pi_\xi -\delta \bm \Xi_\xi = d\bm \Sigma_\xi~.
\ee

We can define the sympletic form 
\be
\bm \omega[ \delta_1\phi, \delta_2 \phi]=\delta_1 \bm \Theta[\delta_2\phi] -\delta_2\bm  \Theta[\delta_1\phi] ~,
\ee
and Noether current  
\footnote{ The inner product between a vector field and a differential form is defined as
\be
\xi \cdot  \bm M \equiv \iota_\xi \bm M =\frac{1}{(n-1)!} \xi^\mu M_{[\mu \nu_2...\nu_n]} dx^{\nu_2} \wedge..  \wedge dx^{\nu_n}, \quad 
\bm M=\frac{1}{n!}M_{[\nu_1 \nu_2...\nu_d]} dx^{\nu_1} \wedge..  \wedge dx^{\nu_n}~.
\ee
}
\be
\bm J_\xi =\bm \Theta[\delta_\xi \phi] -\xi\cdot \bm L-\bm \Xi_\xi~.
\ee
 
It is easy to check that the Noether current is conserved  on-shell
\be
d\bm J_\xi =d\bm\Theta - \mathcal L_\xi \bm L -d\bm\Xi =-\bm E_\phi \delta_\xi \phi \simeq 0~.
\ee
In the above derivation we used the  Cartan identity
\be
\mathcal L_\xi =d \iota_\xi+\iota_\xi d~.
\ee

The conservation of Noether current is expected considering the generalized diffeomorphism invariance.  And we can even further define the Noether charge associated to it
\be
d\bm Q_\xi = \bm J_\xi~.
\ee

We can also calculate the variation of the Noether current 
\be\label{deltaJ}
\delta \bm J_\xi =\bm  \omega[\delta \phi, \delta_\xi \phi] +\delta_\xi \bm \Theta[\delta\phi] 
 -\xi \cdot (d\bm \Theta +\bm E_\phi \delta \phi) -\delta \bm \Xi_\xi
=
\bm  \omega[\delta \phi, \delta_\xi \phi]+ d(\bm \Sigma_\xi+ \xi \cdot \bm \Theta)  ~,
\ee
where we used the relation \eqref{PXS}.

To proceed we need to find a quantity $\bm C_\xi$ such that 
\be\label{CTS}
\delta \bm C_\xi =\xi \cdot\bm \Theta+\bm\Sigma_\xi~,
\ee
and define 
\be
\widetilde{ \bm Q}_\xi =\bm Q_\xi - \bm C_\xi ~.
\ee

Then we can prove that 
\be
\delta d \widetilde{ \bm Q}_\xi = \delta \bm J_\xi - (\xi \cdot\bm  \Theta+\bm \Sigma_\xi) = \bm  \omega~.
\ee
This means that $\widetilde {\bm Q}_\xi  $  is the Hamiltonian (density) generating  the diffeomorphism under $\xi$. 

More precisely, we can define the following quantity which generates the diffeomorphism
\be
\delta H_\xi =\int_{\mathcal C} \bm\omega[\delta\phi, \delta_\xi \phi]
= \int_ {\mathcal C} \Big( \delta \bm J_\xi - d(\bm {\mathcal C}_\xi+ \xi \cdot \bm \Theta)   \Big) 
= \int_{\p{\mathcal C} } \Big(  \delta\bm Q_\xi- \bm{\mathcal C}_\xi - \xi \cdot \bm \Theta    \Big) ~.
\ee
We will always use $\bm \chi$ to denote the integrand, namely  $ \delta H_\xi =\int_{\p{\mathcal C} }  d\bm \chi$.

If we can find $\bm N_\xi[\phi], \bm K[\phi]$   such that 
\footnote{This is possible when the integrability condition is satisfied: 
\be
\int_{\p\mathcal C} \xi \cdot \bm \omega[\delta_1 \phi, \delta_2\phi]=0
\ee
}
\be
\delta \bm N_\xi=\bm \Sigma_\xi, \qquad \delta(\xi\cdot \bm K) =\xi \cdot \bm \Theta[\delta\phi]  
\quad on \quad \p\mathcal C~,
\ee
then  we can integrate in the phase space and define 
\be\label{covHam}
H_\xi  = \int_{\mathcal C}  \bm J_\xi - \int_{\p{\mathcal C}} \Big( \bm N_\xi+ \xi \cdot \bm K    \Big) 
= \int_{\p\mathcal C } \Big( \bm Q_\xi-\bm  N_\xi - \xi \cdot \bm  K    \Big) 
= \int_{\p\mathcal C } \widetilde {\bm Q}_\xi  ~,
\ee
where we used the relation 
\be
\xi\cdot\bm  K=\bm C_\xi-\bm N_\xi~,
\ee
which can be derived from  \eqref{CTS}. 

It is worth mentioning that although various quantities suffer ambiguities in their definition, the integrated charge is well-defined and physical. And if the CS term  is absent, $\bm \Xi, \bm \Pi, \bm \Sigma, \bm C, \bm {\tilde Q}, \bm N$ are not needed, as in the pure Einstein gravity case. 

\subsection{Explicit expressions for TMG}


In this paper, we consider the TMG  which consists of Einstein-Hilbert term (including cosmological  constant term) and CS term. The linearity of the charge enables us to consider them separately. For the pure CS term, we can  make a choice such that $\bm\Sigma_\xi =0$, hence also $\bm N_\xi=0$ \cite{Tachikawa:2006sz}. 
 Some explicit expressions for Einstein-Hilbert term and CS term  are given as follows \cite{Faulkner:2013ica, Cheng:2015nwe}. 

The integrands in the Hamiltonian generating infinitesimal diffeomorphism are
  \beqn
 {\bm \chi}_\xi^{EH} [\delta g ] &=& \delta \bm Q^{EH} _\xi -\xi \cdot \bm \Theta^{EH} 
 \\&=&
 \frac{1} { 16\pi G_N}  \bm \epsilon_{\mu\nu}
  \Big( \xi^\mu \nabla_\sigma \delta g^{\nu\sigma} -\xi^\mu\nabla^\nu \delta g_\sigma^{\; \sigma}
  +\xi_\sigma\nabla^\nu \delta g^{\mu\sigma}
  +\frac12  \delta g_\sigma^{\; \sigma} \nabla^\nu \xi^\mu-\delta g^{\sigma \nu}\nabla_\sigma \xi^\mu
  \Big) ~,
  \qquad\qquad
  \label{EHchi}
\\
{ \bm \chi}_{\xi}^\text{CS}[\delta g]&=& \delta \tilde {\bm Q}_\xi^\text{CS} 
=\frac{1}{32\pi G_N\mu}\Bigg[   2 \delta \Gamma^\nu_{\beta\mu}  \Big(\p_\nu \xi^\mu 
 +\xi^\rho \Gamma^\mu_{\nu\rho}  \Big) 
 + \xi^\alpha \Big(\delta g_{\alpha\beta} R+g_{\alpha\beta} \delta R-4 \delta R_{\alpha\beta} \Big) 
\nonumber\\&&  \qquad\qquad\qquad  \qquad\qquad\qquad  \qquad\qquad 
 -2g^{\mu\nu}\xi^\alpha \Big(\delta g_{\beta\mu} R_{\alpha \nu}-\delta g_{\alpha\mu}R_{\beta\nu} \Big)
 \Bigg] dx^\beta~.
   \label{CSchi}
 \eeqn

The Noether charges are  
  \beqn \label{QEH}
\bm Q_\xi^{EH} &=& -\frac{1}{16\pi G_N}\nabla^\mu \xi^\nu \bm \epsilon_{\mu\nu}~,
\\
\bm Q_\xi^\text{CS} &=& \frac{1}{32\pi G_N\mu}\Bigg[    \Gamma^\nu_{\beta\mu}  \Big(2\p_\nu \xi^\mu 
 +\xi^\rho \Gamma^\mu_{\nu\rho}  \Big) 
 + \xi^\alpha \Big(\delta g_{\alpha\beta} R+g_{\alpha\beta} \delta R-4 \delta R_{\alpha\beta} \Big) 
  \Bigg] dx^\beta~.
\eeqn

 The Noether currents are  
 \beqn
\label{EHJ}
{\bm J}_\xi^\text{EH} &=&  d{\bf Q}_{\xi}^\text{EH}+      2 \xi^\mu E^g_{\mu\nu}{\bm \epsilon}^\nu,\qquad
E^g_{\mu\nu}=\frac{1}{16\pi G_N} (R_{\mu\nu}-\frac{1}{2}g_{\mu\nu}R+\Lambda g_{\mu\nu})~,
\\
\label{CSJ}
 {\bm J}_\xi^\text{CS} &=&  d{\bf Q}_{\xi}^\text{CS}+ \frac{1}{32\pi G_N\mu}    2\xi^\mu C_{\mu\nu} {\bm \epsilon}^\nu~.
\eeqn
Note that these expressions hold \emph{off-shell} and have the form $\bm J=d\bm Q+\xi^a \bm C_a$  with $\bm C_a$ the constraints (equation of motion).

 \section{Linearized equation of motion in TMG from entanglement}\label{eomTMG}
 
 In the previous section, we used the first law of entanglement to derive the holographic dictionary of stress tensor by considering the $R\rightarrow 0$ limit of the interval. It is reasonable to expect that the finite $R$  cases will give  rise to more higher order constraints.  In \cite{Lashkari:2013koa}, the authors  implemented this method for Einstein gravity  and  showed that  all order constraints are equivalent to the linearized Einstein equation. A more elegant approach based on the Wald formalism was used to prove the linearized equation of motion from HEE  in  higher derivative gravities.  In this section, we will adopt a similar strategy and derive  the linearized equation of motion in TMG from HEE. The tool which bridges two sides is the  Wald-Tachikawa formalism which is a generalization of Wald formalism and takes into account the Cherns-Simons contribution, as we have reviewed in the last section. 
 
\subsection{From non-local constraints to integral constraints}

 Before discussing the equation of motion, we first prove two important equations:
  \beqn
\delta E_B^{grav}   &\equiv&   \int_B\bm\chi  =\delta\EV{H}   \label{EB}~,
\\
\delta S_B^{grav}   & \equiv&  \int_{\tilde B}\bm\chi = \delta S_\text{HEE}  \label{SB}~.
 \eeqn

 We specialise to case that the entanglement  interval  $B:  (t=0, x=-R) \rightarrow (t=0, x=R)$ sits on a constant time slice and centers at the origin. We can then use the translation and boost symmetry of the boundary plane to consider  all other intervals. A direct way to consider the boosted interval is also straightforward by using the modular flow generator  \eqref{modflow2} of the boosted interval. 
 
 Let us first consider  the story  on the asymptotic boundary $B$. 
Substituting the metric fluctuations  \eqref{metricFluctuation} into \eqref{EHchi} and \eqref{CSchi}, we get
\be
\int_B \bm\chi^\text{EH}
=\frac{1} { 16\pi G_N}  \frac{2\pi}{R} \int_{-R}^R \Big[  (R^2-x^2) h_{xx} (z,0,x) +\frac{z}{2} (R^2-x^2-z^2) \p_z h_{xx}(z,0,x) \Big] \Big|_{z=0} dx  ~,  \qquad
\ee
and
\beqn
\int_B \bm\chi^\text{CS}
&=&\frac{1} { 16\pi G_N}  \frac{4\pi}{R} \int_{-R}^R \Big[  (R^2-x^2) h_{tx}(z,0,x) +xz^2 (\p_x h_{tx}(z,0,x) -\p_t h_{xx}(z,t,x)|_{t=0})
\nonumber\\&&\qquad
+z(2R^2-2x^2-z^2) \p_z h_{tx}(z,0,x) +\frac12 z^2 (R^2-x^2-z^2) \p_z^2 h_{tx}(z,0,x)  \Big] \Big|_{z=0}dx~. \qquad\qquad
\eeqn

Since we are now interested in the asymptotic boundary, we can further specialise to the near boundary expansion \eqref{metricepxand} of the metric fluctuations.  Substituting it into the above two expressions, we  get their sum
\beqn
 \delta E_B^{grav}  \equiv \int_B\bm\chi
& =&\frac{1} { 8G_N}   \int_{-R}^R  \frac{R^2-x^2}{ R}
 \Bigg[ H_{xx}(0,x)+B_{xx}(0,x) +\frac{H_{tx}(0,x)+3B_{tx}(0,x)}{\mu}
\nonumber \qquad \qquad \\&& \qquad \qquad \qquad \qquad \qquad
 + 2\log z \Big( B_{xx}(0,x)+\frac{1}{ \mu}   B_{tx}(0,x) \Big) 
 \Bigg] dx  ~.
\eeqn
The coefficient of $\log z$ should be zero otherwise the quantity is divergent on the boundary. The resulting requirement is exactly the same as the one we derived in \eqref{Bxxtmu}. As argued before, in the presence of the log mode $B_{ij}$, the Lorentz invariance is only preserved at the chiral point $\mu=1$. Thus, 
\beqn\label{EBequalH}
 \delta E_B^{grav}  \equiv \int_B\bm\chi
& =&\frac{1} { 8G_N}   \int_{-R}^R  \frac{R^2-x^2}{ R}
 \Bigg[ H_{xx}(0,x)+\frac{1  }{\mu}H_{tx}(0,x) +2B_{tx}(0,x) \delta_{\mu,1}
  \Bigg] dx  ~.
\eeqn
It is easy to see that this equations is exactly the same as the expression of modular Hamiltonian \eqref{modH} when the holographic dictionaries \eqref{hol1}, \eqref{hol2} are used. Therefore, 
\be
 \delta E_B^{grav}  =\delta \EV{H}~.
\ee
More generally,  using Lorentz invariance or considering arbitrary boosted intervals, we have the covariant expression of the integrand
\be
\bm \chi|_{\p M} =d\Sigma^\mu T_{\mu\nu}^{grav} \zeta^\nu~.
\ee
Especially, the conservation and traceless of the CFT stress tensor implies that 
\be\label{chibdy}
d\bm \chi|_{\p M} =0~.
\ee
This property ensures that on the boundary we can actually consider arbitrary path which connects the endpoints of $B$ and $ \delta E_B^{grav}  $ is independent of the choice of the path. This is consistent with the requirement that the entanglement entropy should be  a function of the causal development of the interval. 

Now we switch to the story on the geodesic $\tilde B$. This is a   bifurcate   horizon such that $\xi|_{\tilde B}=0$ and $\nabla^a \xi^b|_{\tilde B}= 2\pi \epsilon^{ab}$. 

For the Einstein part, using \eqref{QEH}, we have
\be\label{Qlength}
\int_{\tilde B} \bm Q_\xi^\text{EH} =-\frac{1}{16\pi G_N}\int_{\tilde B} \nabla^a \xi^b \bm \epsilon_{ab}
=\frac{1}{4G} {\sf Length }(\tilde B)~,
\ee
thus
\be
\int_{\tilde B} { \bm \chi}_{\xi}^\text{EH}  =\int_{\tilde B} \delta \bm Q_\xi^\text{EH}
=\frac{1}{4G} \delta{\sf Length }(\tilde B)=\delta S_\text{RT}~.
\ee

While for the  CS part, the infinitesimal charge    \eqref{CSchi} simplifies a lot on the   bifurcate   horizon where the modular flow vanishes
\be
{ \bm \chi}_{\xi}^\text{CS} |_{\tilde B}=\frac{1}{16\pi G_N\mu} \delta \Gamma^\nu_{\beta\mu}  \p_\nu \xi^\mu  dx^\beta~,
\ee
therefore
\be
\int_{\tilde B} { \bm \chi}_{\xi}^\text{CS} |_{\tilde B}=\frac{1}{16\pi G_N\mu}\int_{\tilde B} dx^\beta \delta \Gamma^\nu_{\beta\mu}  \p_\nu \xi^\mu 
=\delta S_\text{CS}~,
\ee
where for the last equality, we observe  that the expression in the middle coincides with \eqref{deltaScs}. 

So we conclude that
\be
 \delta S_B^{grav}  \equiv \int_{\tilde B}\bm\chi
=\int_{\tilde B}\bm\chi^\text{EH}+\int_{\tilde B}\bm\chi^\text{CS}
=\delta S_\text{RT}+\delta S_\text{CS} \equiv\delta S_\text{HEE}~.
\ee

These complete the proof of \eqref{EB} and \eqref{SB}. Especially they hold \emph{off-shell}.

Finally, we also  would like to show 
 \be\label{dchi}
 d\bm\chi =- 2 \xi^\mu    \delta   \mathcal E _{\mu\nu}  {\bm \epsilon}^\nu  ~,
 \ee
 where the equation of motion tensor $  \mathcal E _{\mu\nu}$  is given in \eqref{eomTensor}.
 
To show this, note that when  $\xi$ is a Killing vector of the   AdS background, $\bm\omega[\delta\phi,\delta_\xi\phi]=0$. Together with the choice $\bm\Sigma_\xi=0$ in CS, \eqref{deltaJ} simplifies as $\delta \bm J_\xi=d (\xi\cdot \bm\Theta)$. Therefore
\be\label{dchi2}
d {\bm \chi}_\xi
=d(\delta \bm Q_\xi-\xi\cdot \bm \Theta)
 =d(\delta \bm Q_\xi -  \bm J_\xi )
 =-\xi^a \delta\bm C_a
=- 2 \xi^\mu    \delta    \mathcal E_{\mu\nu}  {\bm \epsilon}^\nu  ~,
\ee
where the off-shell expression of Noether currents \eqref{EHJ} and \eqref{CSJ} are used.

Explicit calculations have also verified that 
\beqn
 d {\bm \chi}_\xi^\text{EH}
&=&
- 2 \xi^\mu    \delta   E^g_{\mu\nu}  {\bm \epsilon}^\nu  ~,
\\
 d {\bm \chi}_\xi^\text{CS}
&=&
  \frac{1}{8\pi G_N \mu} \xi^\mu    \delta   C_{\mu\nu}  {\bm \epsilon}^\nu  ~.
\eeqn
So \eqref{dchi} is indeed true at the linearized level.

 \subsection{From  integral constraints to the linearized equation of motion}
 
 Using identities proven in the last subsection, the first law of entanglement can be rewritten as
 \be
0=\delta S_\text{HEE} - \delta\EV{H} = \delta S_B^{grav}-\delta E_B^{grav}=   \int_{\tilde B}\bm\chi-  \int_{  B}\bm\chi ~.
 \ee
 Since $B$ and $\tilde B$ enclose a  surface $\mathcal C$ such that $\p\mathcal C=B+\tilde B$,    we can  us Stoke's theorem to further rewrite  it as
 \be\label{eomdchi}
 \int_{\p \mathcal C}\bm \chi=\int_{\mathcal C} d\bm \chi =-2 \int_{\mathcal C} \xi^{\mu } \delta\mathcal E_{\mu\nu }\bm\epsilon^\nu=0  ~,
 \ee
 where we used another identity \eqref{dchi} proved in the last subsection.  
    
So far, in the above derivation, we specialize to the case that $B$ is on the constant time slice centering at the origin. Translational invariance and Lorentz invariance on the boundary ensure  that the above equation is valid for arbitrary interval $B$ and its associated $  \mathcal C$. This can also be verified explicitly by re-running the derivation above using the modular flow \eqref{modflow2} of boosted interval. 

The infinitely many constraints \eqref{eomdchi} associated to the infinitely many intervals  enables us to  promote the integrated constraint to a local constraint $\delta  \mathcal E_{\mu\nu}=0$. The details of the argument are almost identical to the pure Einstein gravity case and can be found in  \cite{Faulkner:2013ica}.  However, some care should be taken for TMG.

More specifically, we can first consider the interval $B(R,x_0)$ on the constant time slice and centering at $(x_0,t_0=0)$. Multiplying \eqref{eomdchi} by $R$ and taking derivatives with respect to $R$, we get
\be
\int_{\tilde  B} R \xi_B^t\delta \mathcal E_{tt} \hat r \cdot \bm \epsilon^t
+2\pi R \int_{\mathcal C} \delta \mathcal E_{tt} \bm \epsilon^t=0~.
\ee
Since $\xi_B$ vanishes on $\tilde B$, we have
\be
 \int_{\mathcal C} \delta \mathcal E_{tt} \,\bm \epsilon^t=0 
\ee
for arbitrary $\mathcal C(R,x_0,t_0=0)$. These infinitely many equations imply  \cite{Faulkner:2013ica} the vanishing of the integrand  everywhere, namely $   \mathcal E_{tt}(z,x,t=0) =0, \, \forall z,x$. Translational invariance further guarantees that $   \delta \mathcal E_{tt}(z,x,t ) =0, \, \forall z,x,t$.

The $tt$ component of the linearized equation of motion above can be be written in a covariant form  as  $u^a u^b \delta \mathcal E_{\mu\nu}(x,t ) =0,$ for specific choice of $u^a$ with $a,b\in\{t,x\}$. Lorentz invariance requires that it should hold  for arbitrary reference frame $u^a$, thus yielding
\be\label{eommunu}
\delta\mathcal E_{ab}(z,x,t ) =0, \qquad \forall z,x,t~.
\ee
Finally it remains to show that $ \delta \mathcal E_{zz}=0$ and $ \delta \mathcal E_{zt}=\delta \mathcal E_{zx}=0$. This can be done by appealing to the initial value formulation of gravity and by thinking gravity evolves along radial direction. With this formulation, if the constraints $ \delta \mathcal E_{zz}=\delta \mathcal E_{zt}= \delta \mathcal E_{zx}=0$  is true at $z=0$ and \eqref{eommunu}  holds everywhere, then these constraints hold for all $z$. The vanishing of constraints vanishes at $z=0$ immediately follows from the fact that on the boundary  
\be d {\bm \chi}_\xi |_{\p M}
=- 2 \xi^\mu    \delta    \mathcal E_{\mu\nu}  {\bm \epsilon}^\nu   |_{\p M}=0~,
\ee 
as a consequence of the traceless and conservation of stress tensor.  

More concretely, we can use the   Noether identity linearized about the AdS background, 
\be
\nabla^\mu \delta \mathcal E _{\mu\nu}=0~.
\ee
This is verified explicitly in TMG, especially  $\nabla^\mu \delta C _{\mu\nu}=0$.

Using the vanishing of  other components   \eqref{eommunu}, the most general solution to the above equation is given by 
\be
 \delta \mathcal E _{zt}=z C_t,\qquad  \delta \mathcal E _{zx}=z C_x, \qquad
  \delta \mathcal E _{zz}=  C_z-\frac12 z^2 \p_x C_x+\frac12 z^2 \p_t C_t ~.
\ee
In addition, we also have the physical requirement  \eqref{chibdy}
\be 
0=d {\bm \chi}_\xi |_{\p M}
=- 2 \xi^\mu    \delta    \mathcal E_{\mu\nu}  {\bm \epsilon}^\nu   |_{\p M}
=-\Big(  \zeta_B^\mu C_\mu +\tilde \zeta_B^z C_z\Big) dt\wedge dx   ~,
\ee 
where $\tilde \zeta_B^z=-2\pi t/R$. The above equation holds for arbitrary interval and its associated $\zeta$, thus implying that $C_\mu=C_z=0$. Therefore $ \delta\mathcal E_{zz}= \delta \mathcal E_{zt}= \delta\mathcal E_{zx}=0$.

So, we have derived all the components of the linearized equation of motion in TMG using entropic consideration and covariant phase space formalism.

 \section{Relative entropy }\label{relEnt}
 
 The relative entropy provides a way to  measure  the distance of two states in the Hilbert space. In this  section, we will discuss  the holographic dual of relative entropy when the CFTs suffer  gravitational anomalies on the boundary. 
 
\subsection{Relative entropy in the field theory}
Relative entropy measures the distinguishability between a state $\rho$ and a reference state $\sigma$. It is defined as 
\be
S(\rho||\sigma)=\tr(\rho \log \rho)-\tr(\rho \log \sigma)  ~.
\ee

It can also be written as 
\be
S(\rho||\sigma)=  \Big(\tr(\sigma \log \sigma) -\tr(\rho \log \sigma) \Big) + 
\Big( \tr(\rho \log \rho)-\tr(\sigma \log \sigma)  \Big)
= \Delta \EV{H_\sigma}-\Delta S  ~,
\ee
with $H_\sigma=-\log \sigma$   the modular Hamiltonian   and  $S(\rho)=-\tr(\rho \log \rho)$ the von Neumann entropy of $\rho$. We can also consider the relative entropy associated with the  subsystem $A$ by using  the reduced density matrix $\rho_A,\sigma_A$. 

Relative entropy has two properties:
\begin{itemize}

  \item Positivity: 
  \be\label{posRel}
  S(\rho||\sigma) \ge 0~.
  \ee

   \item Monotonicity:  
 \be\label{MonRel}
 \text{If } \quad A\subset B ,\qquad \text{then } \quad  S(\rho_A||\sigma_A) \le S(\rho_B||\sigma_B)~.
 \ee
   
\end{itemize}

\subsection{Relative entropy in the bulk}

Using the holographic correspondence of various quantities, the relative entropy in CFT can be written in terms of bulk quantities as
\be
S(\rho_{B} ||\sigma_{B}^{vac} )=  \Delta \EV{H_\sigma}-\Delta S
=  \int_{B} \zeta_B^\mu {T}_{\mu\nu}\epsilon^\nu - \Delta S_\text{HEE} (\tilde B ), 
\quad   S_{\text{HEE}}=\frac{1}{4G_N}{\sf Length}+\frac{1}{4G_N\mu} \sf{Twist}  ~,
\ee
where the holographic dictionary of stress tensor \eqref{hol1}, \eqref{hol2} should be used to converted it into a expression in terms of bulk metric.

Our goal in this section is to generalize the holographic dual of  relative entropy in  pure Einstein gravity~\cite{Faulkner:2013ica}  to TMG and  prove that 
\be\label{holRE}
S(\rho_{B} ||\sigma_{B}^{vac})
=H_{\xi_B}(M) -H_{\xi_B} (\text{AdS})~,
\ee
where $H_\xi$ is the Hamiltonian generating the diffeomorphism in phase space given by \eqref{covHam}. For general asymptotically AdS spacetime $M$, there are no Killing vectors.  But since the relative entropy finally only involves $B$ and $\tilde B$ in the bulk, we can find a  vector $\xi$ which shares the same behavior as the Killing vector in pure AdS when they approaches $B$ and $\tilde B$. More specifically, we have the following requirement \cite{Faulkner:2013ica} 
\beqn
\xi^\mu|_{B} &=& \zeta^\mu_{B} ~, \\
\nabla_{(\mu} \xi_{\nu)}  |_{z \rightarrow 0} &= &\mathcal O(1 ) ~,  \\
 \nabla^{[\mu }\xi^{\nu]}|_{  {\tilde B}} &= &2\pi \epsilon^{\mu\nu} ~,\\
 \xi |_{  {\tilde B}} &=&0  ~,
 \eeqn
where $\epsilon^{\mu\nu}=  \tilde n^\mu   n^\nu -  \tilde n^\nu  n^\mu$ is the binormal  \eqref{binormal} and $\zeta_B$ is the conformal Killing vector  \eqref{boundaryflow} on the boundary.  The vector $\xi$ always exists and  an explicit construction can be found in \cite{Faulkner:2013ica}. Since we only impose constraints near $B$ and $\tilde B$, the vector $\xi$ is actually not unique but   the physical $H_\xi$ is  always unambiguous.

The construction of $\xi$ is almost identical to the pure Einstein gravity case, but the new ingredient comes from the extra    contribution from CS term in $H_\xi$.  Four the pure CS term in 3D, as we said, we can make a choice such that $\bm\Sigma_\xi =0$, hence also $\bm N_\xi=0$.  Then it is easy to see that for both Einstein-Hilbert term and CS term, the Hamiltonian is given by  
\be
H_\xi= \int_{\mathcal C} \bm   J _\xi - \int_{\p{\mathcal C}}  \xi \cdot \bm K 
=\int_{\p\mathcal C} \Big(\bm  Q_\xi  - \xi \cdot \bm K    \Big) ~.
\ee
More specifically, the  two contributions are 
\beqn
H^\text{EH}_\xi  &=& \int_{\mathcal C} \bm   J^\text{EH}_\xi - \int_{\p{\mathcal C}}  \xi \cdot \bm K ^{EH}    
= \int_{\p{\mathcal C} } \Big(\bm  Q^{EH}_\xi  - \xi \cdot \bm K^{EH}    \Big) ~.
\\
H_\xi^\text{CS} &=& \int_{\p{\mathcal C} } \Big(\bm  Q^\text{CS}_\xi  - \xi \cdot \bm K^\text{CS}    \Big) =\int_{\p\mathcal C} \tilde{ \bm Q}_\xi ^\text{CS}
\eeqn
For pure Einstein gravity, $\bm K^{EH }$ is just the standard Gibbons-Hawking term.  More generally, it can be identified through holographic renormalization. \footnote{For the CS term, only $\tilde{\bm Q}_\xi $ is the well-defiend Noether charge. So the associated $\tilde {\bm K}^\text{CS}=0$.  This is consistent with  the fact that for the CS term no extra boundary term is needed for a well-defined  variation~\cite{Kraus:2005zm}.}

So the final Hamiltonian is 
\be
H_\xi=H_\xi ^{EH}+H_\xi ^\text{CS}=\int_{\p\mathcal C}\Big[\tilde { \bm Q}_\xi ^\text{CS} +{ \bm Q}_\xi ^{EH }-\xi\cdot \bm K^{EH } \Big]~.
\ee

In order to prove \eqref{holRE}, we need to show that 
\beqn \label{RelonB}
\Delta  \int_B  \Big[\tilde { \bm Q}_\xi ^\text{CS} +{ \bm Q}_\xi ^{EH }-\xi\cdot \bm K^{EH } \Big]  &=&
  \int_{B} \zeta_B^\mu {T}_{\mu\nu}\epsilon^\nu ~,
\\
\label{RelonBt}
\Delta  \int_{\tilde B}  \Big[\tilde { \bm Q}_\xi ^\text{CS} +{ \bm Q}_\xi ^{EH }-\xi\cdot \bm K^{EH }  \Big] &=& 
\frac{1}{4G_N}{\Delta\sf Length}+\frac{1}{4G_N\mu} \Delta\sf{Twist}  ~.
\eeqn

Let us start with \eqref{RelonBt}. 
Since $\xi$ vanishes on $\tilde B$, using \eqref{Qlength}  we easily see that 
\be 
\Delta  \int_{\tilde B}  \Big[ { \bm Q}_\xi ^{EH }-\xi\cdot \bm K^{EH }  \Big] 
=\Delta  \int_{\tilde B}    { \bm Q}_\xi ^{EH }  
=\frac{1}{4G_N}{\Delta\sf Length} ~.
\ee

Thus,  it boils down to show 
 \be\label{finiteDiff}
  \Delta  \int_{\tilde B}    { \tilde{\bm Q}}_\xi ^{CS }  
=\frac{1}{4G_N \mu}{\Delta\sf Twist}~.
 \ee

The infinitesimal counterpart of this expression is 
 \be
\delta S_\text{CS}= \frac{1}{4G_N \mu}{\delta\sf Twist}
 =\delta \int_{\tilde B}    { \tilde{\bm Q}}_\xi ^{CS }  
=\int_{\tilde B}      \bm \chi^{CS }  
=\frac{1}{16\pi G_N\mu}   \int dx^\beta \delta \Gamma^\nu_{\beta\mu}   \p_\nu \xi^\mu ~,
 \ee
 where we used  \eqref{CSchi} and the fact that $\xi$ vanishes on $\tilde B$. Compared with the expression derived from HEE proposal \eqref{deltaScs}, we  see that the above equation is indeed true.  The  equation \eqref{finiteDiff} can then be proved by choosing an   path  which connects the pure AdS and $M$ in the phase space of metric  and then integrating along the path. 

To show \eqref{RelonB}, wen can similarly consider its infinitesimal version first
\be
\delta  \int_B  \Big[\tilde { \bm Q}_\xi ^\text{RT} +{ \bm Q}_\xi ^{EH }-\xi\cdot \bm K^{EH } \Big]  =
\int_B \bm \chi =
  \int_{B} \zeta_B^\mu  \delta{T}_{\mu\nu}\epsilon^\nu~.
\ee 
This indeed holds, as we have shown in \eqref{EBequalH} in the last section. Again, we can integrate this infinitesimal expression over an arbitrary path in phase space from AdS to desired spacetime $M$. This resulting expression is exactly  \eqref{RelonB}. The path independence of the expression also suggests the existence of  $\bm K^{EH}$.

Therefore, we prove that the holographic dual of relative entropy in CFTs with gravitational anomalies is given by \eqref{holRE}, the difference of quasi-local energy in the corresponding entanglement wedge.

\subsection{Implications}

In the last subsection, we proved that the relative entropy in CFTs with gravitational anomalies is holographically dual to the vacuum-subtracted energy in the dual spacetime. This holographic correspondence enables us to translate the quantum information inequalities \eqref{posRel},\eqref{MonRel} in CFTs into new positive energy theorems  of TMG in asymptotically AdS spacetime.

More specifically, the positivity  \eqref{posRel} of relative entropy implies that the vacuum-subtracted energy associated to  every interval $B$, or equivalently  every  entanglement wedge, is positive definite. 
Furthermore, the monotonicity  \eqref{MonRel}  of relative entropy implies that the vacuum-subtracted energy should increase when the size of $B$ and the associated entanglement wedge becomes larger.  

The global positive energy theorem of TMG was discussed in \cite{Deser:2009ki,Sezgin:2009dj} by generalizing the  spinor techniques in \cite{Witten:1981mf}, although many issues are quite obscure there.  The global positive energy theorem is supposed to correspond  to a special case of our more general positive energy theorems when $B$ is taken as the global spatial slice of the boundary and  $\xi_B$ coincides with the time translation on the boundary. The positive energy theorems in this paper are much stronger; there are infinitely many positive energy theorems associated to infinite subregions on the boundary. The underlying ground of our positive energy theorems is also quite different. Traditionally the positive energy theorem follows from some types of energy conditions. While here our derivations are based on the holographic principle. Therefore, as long as the spacetime admits a CFT dual, then our positive energy theorem should hold. On the other hand, this infinite set of positive energy theorem also provides  a criterion to test whether an arbitrarily given spacetime arises from a consistent UV quantum gravity whose low energy description is TMG. Any spacetime geometry or theory violating these positive energy theorems is thus in the swampland.  

Interestingly, it has been argued in \cite{ Li:2008dq,Maloney:2009ck} that the TMG itself is   unstable/inconsistent generically  due to  the negative energy of either massive gravitons or BTZ black holes.  The only possible UV completable TMG is chiral gravity with $\mu\ell=1$.  Therefore, it would be very interesting and significant  if we can use the positive energy theorems, enforced by holography and quantum information inequalities, to show the UV inconsistency of non-chiral TMG. This is especially natural considering that relative entropy takes in to account the  higher order perturbations, which are relevant in the analysis of inconsistency in TMG \cite{Maloney:2009ck,Li:2008dq}. We leave this interesting question for the future.

 \section{Conclusion} \label{conclu}

In this paper, we explored some aspects of entanglement in the context of AdS$_3$/CFT$_2$  in the presence of  gravitational anomalies. We derived the holographic dictionary of stress tensor based on the first law of entanglement and Lorentz invariance. Especially the dictionary for TMG at  the chiral point is also found and the logarithmic mode appears automatically.  
 
 We further use the first law of entanglement to derive  the linearized  equation of motion in TMG. Therefore not only do  the Einstein and higher derivative   gravitation, but also the anomalous gravitation, emerge from entanglement. This provides further evidence that the entanglement may be  the underlying way of organising our spacetime. 
 
 Finally, we found the holographic dual of relative entropy when the boundary CFTs suffer gravitational anomalies. The positivity of relative entropy helps establish   generalized positive energy theorems for TMG in AdS.  
 
In this paper, for simplicity we use the Poincare AdS as the background. But we can  also  consider the general zero mode BTZ background and make similar discussions. Especially the modular flow generator of any subregion in BTZ background is given in appendix~\eqref{appBTZ}. With this, the linearized equation of motion about BTZ background in TMG can also be derived straightforwardly.  These discussions can also be performed similarly  in other holographic setups \cite{AJSZ}.
 
There are many other interesting problems worth further studying.  Some of them are  as follows.

 \subsection*{\bf \emph{Higher dimensional generalizations}}
 
 In this paper, for simplicity we focus exclusively on AdS$_3$/CFT$_2$. It is interesting to generalize our discussions to higher dimension.   The pure/mixed gravitational CS terms exist  in   odd dimension $d=2k+1$. Via the anomaly inflow mechanism \cite{Callan:1984sa},  it yields the anomalies on the boundary with even dimension $d=2k$, more specifically gravitational anomalies in $d=4k+2 $  and  the mixed  gauge-gravitational   anomalies  in $d=2k, k \in \mathbb Z_{> 1}$.  
The HEE in the presence of pure/mixed gravitational  CS terms in higher dimensions has been considered in~\cite{Azeyanagi:2015uoa}. Using their proposal  and the first law of entanglement, many discussions in this paper can be generalized straightforwardly, especially  the derivation of holographic dictionary of stress tensor and the linearized equation of motion.

  \subsection*{\bf \emph{Equation of motion beyond the linearized order}}

 In this paper, we derived the linearized equation of motion of bulk gravity in the presence of CS term. For AdS/CFT with Einstein gravity and higher derivative gravity, the equation of motion beyond linearized order were fulfilled in \cite{Faulkner:2017tkh,Haehl:2017sot}. It is interesting and important to see whether some ideas and methods there can be extended to incorporate the CS term and gravitational anomalies.

\subsection*{\bf \emph{Implications and constraints on bulk gravity from relative entropy} }

 In this paper, we found a notion of quasi-local energy in gravitational systems with the  CS term  and established its positivity using the quantum information inequality, generalising the  pure Einstein gravity case~ \cite{Lashkari:2016idm}.    It will be interesting to further explore the implications of these generalized positive energy theorems and understand their constraints on theory and spacetime geometry.
 
 For  pure Einstein gravity, it was shown that these  positive energy theorems   are related to many other quantities, including  canonical energy, Fisher information, etc.  In particular for pure Einstein gravity in AdS$_3$,  the interplay between energy conditions and information equalities   was studied in \cite{Lashkari:2014kda}. It would be of interest to extend their discussions to account for the gravitational anomalies and understand the energy condition in TMG.
   

  It will be even more interesting if we can use the quantum information inequalities to gain insight on  the instability and non-unitarity problems for TMG itself. By studying the perturbations around AdS carefully, one may be able to see their (in)compatibility with positive energy theorems, and thus test whether the theory is in the swampland or not. If this analysis can be done explicitly for TMG, in the favourable case, it would be a quite strong support of chiral gravity as  a UV consistent theory. And the global positive theorem in  chiral gravity  can be simply proved since it is just a special case of the general positive energy theorems  here.

  \vspace{0.5cm}
  
  The interplay between entanglement and anomaly is   revealing some profound aspects of quantum field theory and holography. We hope to report some   progress along these directions in the future.

   
\section*{Acknowledgement}

  The author would like to thank L. Hung for correspondence and W. Song for useful comments on the draft. The author also   thank L. Apolo, W. Song and Y. Zhong for collaboration on related problem. This work is supported by Swiss National Science Foundation.

 \appendix


 \section{Modular flow Killing vectors in arbitrary zero-mode background}  \label{appRindler} \label{modFlowandRindler}
 
In this appendix, we will use the Rindler method  \cite{Casini:2011kv,Castro:2015csg,Jiang:2017ecm}  to calculate the HEE. The Rindler transformation maps the causal development of a subregion into an infinite Rindler spacetime. Accordingly, the entanglement entropy becomes the thermal entropy in the Rindler spacetime due to the unitarity of the Rindler transformation. We will use this method to compute the HEE for subregion in Poincare AdS. 

Another main goal in this appendix is to calculate the modular flow generator  in the BTZ background associated to arbitrary interval. For Poincare AdS, we compute the modular flow Killing vector using Rindler method. More generally, the modular flow Killing vectors are computed through some physical requirements. 
 
\subsection{Poincare AdS}\label{appPAdS}

We consider the TMG in the Poincare  AdS 
\be
 ds^2=\frac{dr^2}{4r^2}+2r du dv~.
 \ee  
And the subregion of our interest is an arbitrary interval on the boundary
    \be\label{EEinterval}
 - (\frac{l_u}{2},\frac{l_v}{2}) \rightarrow (\frac{l_u}{2},\frac{l_v}{2})~.
   \ee

  The isometry generators  of Poincare AdS are
 \beqn\label{Lm}
 L_{-1}=\p_u, \qquad L_0= -u \p_u + r \p_r , \qquad   L_1=u^2\p_u -\frac{1}{2r} \p_v- 2 r u \p_r ~,
&& \\
\label{Lbarm}
\bar  L_{-1}=\p_v, \qquad \bar  L_0= - v \p_v+ r \p_r , \qquad \bar  L_1=v^2\p_v -\frac{1}{2r} \p_u- 2 r v \p_r ~.
 \eeqn
 They form the isometry algebra SL(2,$\mathbb R$)$\times$SL(2,$\mathbb R$), satisfying 
 \be
 [L_m,L_n]=(m-n)L_{m+n}, \qquad  [\bar L_m, \bar L_n]=(m-n)\bar L_{m+n}, \qquad  [L_m, \bar L_n]=0 ~.
 \ee

 We would like to find a Rindler coordinate transformation which maps the Poincare AdS
 to a Rindler spacetime which is thermal. It turns out one possible choice of the Rinlder spcetime is 
 \be
{ \tilde ds}^2=\frac{d\tilde r^2}{4(\tilde r^2-1)}+2\tilde r d\tilde u d\tilde v + d \tilde u^2 +d \tilde v^2 ~.
 \ee
 This spacetime has horizon at $\tilde r=1$. 
 The map is possible   because both of them are  locally the same with constant curvature. 
 
 To obtain the coordinate transformation, we consider 
 \beqn
 \p_{\tilde u} &=&    - \frac{2}{l_u}  L_1 +\frac{l_u}{2}  L_{-1} ~,  \\
  \p_{\tilde v} &=&     \frac{2}{l_v}  \bar L_1 - \frac{l_v}{2}   \bar L_{-1} ~.
 \eeqn
 
We can then immediately verify that
 \be
  \p_{\tilde u} \cdot  \p_{\tilde u}= \p_{\tilde v}\cdot  \p_{\tilde v}=1 ~.  
 \ee
We can also  obtain $\tilde r(r,u,v)$ through $  \p_{\tilde u} \cdot  \p_{\tilde v}=\tilde r$.  
 
 At the same time, we can obtain $ \p_{\tilde r}$ by requiring 
 \be
 \p_{\tilde r} \cdot  \p_{\tilde u}= \p_{\tilde r} \cdot  \p_{\tilde v}=0~, 
  \qquad  \p_{\tilde r} \cdot  \p_{\tilde r} = \frac{1}{4(\tilde r^2-1)}~.
 \ee

We can consider the matrix $ J=(\p_{\tilde r }, \p_{\tilde u},\p_{\tilde v})$. The inverse $J^{-1}$ is exactly the Jacobian matrix of the coordinate transformation from   $(\tilde r, \tilde u,\tilde v)$ to $(r, u, v)$. 

Explicitly, we find that 
\beqn
\tilde  r &=&- \frac{4 \left(r^2 u^2 l_v^2-(2 r u v+1)^2\right)-r^2
   l_u^2 \left(l_v^2-4 v^2\right)}{4 r l_u l_v} ~,
   \\
  \tilde  u&=& 
  \frac{1}{4} \log \left(\frac{\left(r \left(l_u+2
   u\right) \left(l_v-2 v\right)-2\right) \left(r
   \left(l_u+2 u\right) \left(l_v+2
   v\right)+2\right)}{\left(r \left(l_u-2 u\right)
   \left(l_v-2 v\right)+2\right) \left(r \left(l_u-2
   u\right) \left(l_v+2 v\right)-2\right)}\right)~,
     \\
  \tilde  v&=&  
  \frac{1}{4} \log \left(\frac{\left(r \left(l_u-2
   u\right) \left(l_v+2 v\right)-2\right) \left(r
   \left(l_u+2 u\right) \left(l_v+2
   v\right)+2\right)}{\left(r \left(l_u-2 u\right)
   \left(l_v-2 v\right)+2\right) \left(r \left(l_u+2
   u\right) \left(l_v-2 v\right)-2\right)}\right)~.
\eeqn
 
On the boundary, the coordinate transformation reduces to 
\beqn
  \tilde  u&=&   \text{arctanh } \frac{2u}{l_u}  \label{bdytsf} ~,  \\
  \tilde  v &=& \text{arctanh }\frac{2v}{l_v}    ~,
\eeqn
 which implies a thermal circle
 \be
( \tilde u, \tilde v) \sim  ( \tilde u + i  \tilde  \beta_{\tilde u} , \tilde v+ i  \tilde  \beta_{\tilde u} ) ~,
 \ee
 with $ \tilde  \beta_{\tilde u} =\tilde  \beta_{\tilde u} = \pi $.
 
The modular flow  is given by
 \beqn
\xi&=& \tilde  \beta_{\tilde u}  \partial_{\tilde u}+ \tilde  \beta_{\tilde v}  \partial_{\tilde v}
 \\ &=&\pi  \Big(  - \frac{2}{l_u}  L_1 +\frac{l_u}{2}  L_{-1} + \frac{2}{l_v}  \bar L_1 - \frac{l_v}{2}   \bar L_{-1} \Big)
 \\&=& 4 \pi r \Big(\frac{u}{l_u}-\frac{v}{l_v}\Big)\p_r +\frac{\pi}{2} \Big( l_u -\frac{2}{r l_v }-\frac{4u^2}{l_v} \Big) \p_u
-\frac{\pi}{2} \Big( l_v -\frac{2}{r l_u }-\frac{4v^2}{l_u} \Big) \p_v  ~.
 \eeqn
 
We can then calculate the thermal entropy of the Rindler spacetime which is exactly the HEE in the old spacetime. 
 
  We need to regulate the interval \eqref{EEinterval} and consider 
  \be
  -(\frac{l_v}{2}-\epsilon_v, \frac{l_u}{2}-\epsilon_u) \rightarrow (\frac{l_v}{2}-\epsilon_v, \frac{l_u}{2}-\epsilon_u) ~.
  \ee 
 From the coordinate transformation \eqref{bdytsf}, we obtain
 \be
 \Delta\tilde u=2 \text{arctanh } \frac{2( \frac{l_u}{2}-\epsilon_u )}{l_u}
 =  \log \frac{l_u}{\epsilon_u}+\mathcal{O}(\epsilon_u)~,
 \ee
 and similarly  $  \Delta\tilde v =  \log \frac{l_v}{\epsilon_v}+\mathcal{O}(\epsilon_v)$.
 
The thermal entropy corresponding to Einstein gravity part is just the Bekenstein-Hawking entropy, given by the length of the horizon  $\tilde \Sigma$
 \be
 S_\text{RT}= S_\text{Rindler thermal  entropy }=\frac{\Delta \tilde u+\Delta \tilde v}{ 4G_N}=\frac{1}{ 4G_N}\log \frac{l_u l_v}{\epsilon_u\epsilon_v}~.
 \ee

  The  thermal entropy  arising from CS term  is \cite{Tachikawa:2006sz}
 \be
 S_\text{CS}
 =\frac{1}{8G\mu} \int_{\tilde \Sigma} \tilde \epsilon_\alpha{}^\beta \tilde \Gamma^\alpha_{\beta\sigma} d\tilde  x^\sigma
 =\frac{1}{4G\mu}    \Big( \Delta\tilde u - \Delta\tilde v \Big)~.
 \ee

 So the whole HEE is given by 
 \be
 S_{\text{HEE}}=S_\text{RT}+S_\text{CS}=\frac{1}{ 4G_N}\log \frac{l_u l_v}{4\epsilon^2 }
 +\frac{1}{4G_N\mu}   \log \frac{l_u  }{l_v}
 =\frac{1}{ 2G_N}\log \frac{ R}{\epsilon  }+\frac{\kappa}{2G_N\mu}   ~.
 \ee
 where $l_u=2Re^{2\kappa}, l_v=2Re^{-2\kappa}$.
 
  \subsection{BTZ}\label{appBTZ}

More generally the background geometry is given by  BTZ black hole. In our coordinate system, it takes the form 
\be
 ds^2=G_{\mu\nu} dx^\mu dx^\nu=\frac{dr^2}{4r^2}+\Big(2r+ \frac{U  V }{2r} \Big) du dv+\Big( U  du^2+V dv^2\Big) ~.
 \ee
The modular flow generator in this background,  in principle, can also be calculated through Rindler transformation. But this would be quite complicated. Instead, we can find it through the following tricks. 
 
 First of all, the modular flow generator is a Killing vector, hence satisfying the   Killing equation $\mathcal L_\xi G_{\mu\nu}=0$. Then we know that the modular flow should vanishes at the endpoints of the interval. So we can impose the following conditions  
 \be
 \xi^\mu(r=\infty, u=\pm l_u/2 ,v=\pm l_v/2)=0~.
 \ee
After a long computation, one can find   the following modular flow generator
%
%
     \hspace*{-3cm}
   
\resizebox{.8\textwidth}{!}{  \vbox{
 \beqn  \label{BTZmodFlow}
\xi^r&=&
 \frac{2 \pi  r \left(\sinh \left(2 u \sqrt{U}\right) \sinh \left(\sqrt{V}
   l_v\right)-\sinh \left(2 v \sqrt{V}\right) \sinh \left(\sqrt{U}
   l_u\right)\right)}{\sqrt{U} \sqrt{V} l_u l_v}~,
  \\
\xi^u&=&
   \frac{\pi  \sinh \left(\sqrt{V}
   l_v\right) \left(\left(U V-4 r^2\right) \cosh \left(\sqrt{U} l_u\right)+\left(4 r^2+U
   V\right) \cosh \left(2 u \sqrt{U}\right)\right)+4 \pi  r \sqrt{U} \sqrt{V} \cosh
   \left(2 v \sqrt{V}\right) \sinh \left(\sqrt{U} l_u\right)}{U \sqrt{V} l_u l_v \left(U
   V-4 r^2\right)}~,
 \qquad\qquad   \\
\xi^v &=&
   \frac{\pi  \sinh \left(\sqrt{U} l_u\right) \left(\left(4 r^2-U
   V\right) \cosh \left(\sqrt{V} l_v\right)-\left(4 r^2+U V\right) \cosh \left(2 v
   \sqrt{V}\right)\right)-4 \pi  r \sqrt{U} \sqrt{V} \cosh \left(2 u \sqrt{U}\right)
   \sinh \left(\sqrt{V} l_v\right)}{\sqrt{U} V l_u l_v \left(U V-4 r^2\right)}~.
 \eeqn
 }  }
 
\noindent On the boundary $r\rightarrow \infty$, it reduces to 
\beqn 
\zeta &=&
 \frac{\pi  \sinh \left(l_v \sqrt{V}\right) \left(\cosh \left(l_u
   \sqrt{U}\right)-\cosh \left(2 u \sqrt{U}\right)\right)}{l_u l_v U
   \sqrt{V}} \p_u
  \\&& +
   \frac{\pi  \sinh \left(l_u \sqrt{U}\right) \left(\cosh \left(2 v
   \sqrt{V}\right)-\cosh \left(l_v \sqrt{V}\right)\right)}{l_u l_v
   \sqrt{U} V}   \p_v~.
\eeqn
 In particular, one can verify that in the  limit $U=V=0$, they reduce to the known result \eqref{modflow2}. 
 
 From the Killing vector, we can find the Killing horizon:
 \be
 \mathcal H: \qquad \xi\cdot \xi=0~.
 \ee
 It has two branches: 
  \beqn
  \mathcal H_\pm:\qquad r&=& \frac12 \sqrt{U} \sqrt{V}
   \coth \Big((\frac{l_u}{2}\mp u)\sqrt{U}  \Big) \coth \Big((\frac{l_v}{2} \pm v)\sqrt{V}  \Big)  ~.
   \eeqn 
On every Killing horizon, we can define the surface gravity  there
   \be
\xi^\alpha \nabla_\alpha \xi_\beta =\kappa  \xi_\beta~,
 \ee
 or more simply
  \be
 \kappa^2= -\frac{1}{2}( \nabla_\alpha \xi_\beta)( \nabla^\alpha \xi^\beta)~.
 \ee 
 For the two Killing horizons in our problem, the surface gravities are given by 
 \beqn
   \mathcal H_\pm:\qquad 
\kappa_\pm
=\pm 2\pi  \frac{  \sinh  \left(\sqrt{U} l_u\right) \sinh  \left(\sqrt{V} l_v\right)}{ \sqrt{U }\sqrt{V}  l_u  l_v }~.
 \eeqn
We will use the positive one to define  the surface gravity associated to the Killing vector $\xi$, namely $\kappa_\xi =\kappa _+>0$. Especially in the limit $U,V=0$, $\kappa_\xi=\kappa_+=2\pi$ which is exactly the value we used before. 
 
The two Killing horizons $   \mathcal H_\pm $ correspond to the lower and upper boundary of  the entanglement wedge.  The intersection of two Killing horizons is the  bifurcate  horizon which is also the RT-HRT surface, more precisely the geodesic here. The modular flow vanishes there
 \be
 \tilde B:\qquad \xi=0~.
 \ee
 The position of  $\tilde B$ is determined by 
 \beqn
 \cosh \left(2 u \sqrt{U}\right) &=&\frac{\left(4 r^2+U V\right) \cosh
   \left(\sqrt{U} l_u\right)-4 r \sqrt{U} \sqrt{V} \sinh \left(\sqrt{U} l_u\right) \coth
   \left(\sqrt{V} l_v\right)}{4 r^2-U V} ~,
   \qquad    \qquad \\
   \cosh \left(2 v \sqrt{V}\right) &=& \frac{\left(4
   r^2+U V\right) \cosh \left(\sqrt{V} l_v\right)-4 r \sqrt{U} \sqrt{V} \coth
   \left(\sqrt{U} l_u\right) \sinh \left(\sqrt{V} l_v\right)}{4 r^2-U V} ~.
    \qquad    \qquad 
 \eeqn
 The turning point of the geodesic, namely the deepest point in the radial direction,  is given by 
 \be
 r_*=\frac12   \coth   \left(\frac12\sqrt{U} l_u\right)  \coth   \left(\frac12\sqrt{V} l_v\right), \qquad u_*=v_*=0 ~.
 \ee
 
With the explicit coordinate of geodesic, we can  now calculate the geodesic length
 \beqn
L &=&2\int_{r_*}^{1/\epsilon}  dr \sqrt{G_{\mu\nu} \frac{dx^\mu}{dr}\frac{dx^\nu}{dr}}
\nonumber\\&=&
2\int _{r_*}^{1/\epsilon}  dr
  \frac{4 r^2-U V}{2 r }
  \Bigg( \sqrt{2 r-\sqrt{U} \sqrt{V} \tanh  \frac{l_u
   \sqrt{U}}{2}  \tanh  \frac{l_v \sqrt{V}}{2} }
    \sqrt{2
   r-\sqrt{U} \sqrt{V} \coth  \frac{l_v \sqrt{U}}{2}  \coth
    \frac{l_v \sqrt{V}}{l_u2} }  
\nonumber   \\&& \qquad\qquad \times 
    \sqrt{2 r-\sqrt{U} \sqrt{V} \tanh
    \frac{l_u \sqrt{U}}{2}  \coth  \frac{l_v
   \sqrt{V}}{2} } \sqrt{2 r-\sqrt{U} \sqrt{V} \coth  \frac{l_u
   \sqrt{U}}{2}  \tanh  \frac{l_v \sqrt{V}}{2} } \Bigg)^{-1}
\nonumber  \\&=& \log \frac{2 \sinh(l_u \sqrt{U})  \sinh(l_v  \sqrt{V}) }{\sqrt{UV}\epsilon}  +\mathcal{O}(\epsilon) ~.
 \eeqn
 This is also verified numerically.

For the interval  \eqref{intervalconstT} on the constant time slice, $l_u=l_v=2R$. Thus 
\be
L= \log \frac{2 \sinh(2R \sqrt{U})  \sinh(2R  \sqrt{V}) }{\sqrt{UV}\epsilon} ~.
\ee
According to the RT-HRT proposal, we have
\be
S_\text{RT}=\frac{1}{4G} L=\frac{1}{4G}  \log \frac{2 \sinh(2R \sqrt{U})  \sinh(2R  \sqrt{V}) }{\sqrt{UV}\epsilon} ~.
\ee
This is exactly the same as \eqref{EEnewcoord} once we identify the UV/IR regulator $\epsilon$ and $\varepsilon_\chi$ properly. This also justifies our general modular flow generator. 

The complete result of HEE in TMG also includes the contribution from CS term which can be computed according to the proposal \eqref{HEEproposal}. Instead, we will calculate it by integration in the phase space in the next appendix.

  
%

\section{EE via integration in phase space} \label{EEphase}

In this appendix, we provide an approach to calculate EE by integrating the (infinitesimal variation of) Noether charge in phase space. This is supposed to be a quite universal way and here we use the CS Noether charge as an example to illustrate this. 

In the main text, we worked out the variation of EE   
 \be
\delta S_\text{CS}=\frac{1}{8G_N \mu }\frac{1}{\kappa_\xi} \int_ {\tilde B} dx^\rho\delta\Gamma_{\rho\sigma}^\nu   \partial_\nu\xi^\sigma  ~.
 \ee
 This is a variation in the phase space. So it is possible to obtain the charge through phase space integration. Since the modular flow is only known for BTZ black hole background~\footnote{Note that the Killing vector is now a function of $(U,V)$ and is given in \eqref{BTZmodFlow}. },
  we restrict ourselves to the zero mode phase space which is parametrized by constant $U$ and $V$. The fluctuations is thus $\delta U, \delta V$.

From the BTZ metric
 \be
 ds^2=G_{\mu\nu} dx^\mu dx^\nu=\frac{dr^2}{4r^2}+\Big(2r+ \frac{U  V }{2r} \Big) du dv+\Big( U  du^2+V dv^2\Big)~,
 \ee
we can find the variation of metric 
 \be
 \delta G_{\mu\nu}  =\frac{\partial G_{\mu\nu}  }{\partial U}\delta U
 +\frac{\partial G_{\mu\nu}  }{\partial V}\delta V~,
 \ee
as well as the variation of connection 
  \be
\delta \Gamma=\frac{\partial \Gamma}{\partial U}\delta U+\frac{\partial \Gamma}{\partial V}\delta V~.
 \ee
Collecting all the ingredients and performing some manipulations, the variation of EE is found to be
 \beqn
  \delta S_\text{CS} &=&\frac{2}{8G_N \mu } \int_{r_*}^\infty  dr\Big( \frac{ 2rV(32 X^2+(4r^2-UV)^2\text{csch}^2(l_v\sqrt{V}))}{(4r^2-UV)^3 X} \delta U
\\&&\qquad\qquad
 - \frac{  2rU(32 X^2+(4r^2-UV)^2\text{csch}^2(l_u\sqrt{U}))}{(4r^2-UV)^3 X} \delta V
   \Big) ~,
 \eeqn
 where
 \beqn
 X&=& \sqrt{  r- \frac12 \sqrt{U} \sqrt{V} \tanh  \frac{l_u
   \sqrt{U}}{2}  \tanh  \frac{l_v \sqrt{V}}{2} }
    \sqrt{ 
   r-  \frac12   \sqrt{U} \sqrt{V} \coth  \frac{l_v \sqrt{U}}{2}  \coth
    \frac{l_v \sqrt{V}}{l_u2} }  
\nonumber\\ && \times   
    \sqrt{  r-  \frac12   \sqrt{U} \sqrt{V} \tanh
    \frac{l_u \sqrt{U}}{2}  \coth  \frac{l_v
   \sqrt{V}}{2} } \sqrt{  r-  \frac12   \sqrt{U} \sqrt{V} \coth  \frac{l_u
   \sqrt{U}}{2}  \tanh  \frac{l_v \sqrt{V}}{2} }   ~.  \qquad\qquad
 \eeqn

Performing the integration, one finds that 
 \beqn
 \delta S &=&\frac{1}{4G_N \mu } \Big( \frac{l_u\sqrt{U} \coth(l_u\sqrt{U})-1}{2U}\delta U-\frac{l_v\sqrt{V} \coth(l_v\sqrt{V})-1}{2V}\delta V \Big)
\\
&=&
 \frac{1}{4G_N \mu} \delta   \log \Bigg( \frac{ \sqrt V \sinh  (l_u\sqrt U)}{\sqrt U  \sinh  (l_v\sqrt V)} \Bigg) ~.
 \eeqn
 Then  we can integrate in the phase space parametrized by $U,V$ and find
 \be
 S_\text{CS}=\frac{1}{4G_N \mu}  \log  \Bigg( \frac{ \sqrt V \sinh  (l_u\sqrt U)}{\sqrt U  \sinh  (l_v\sqrt V)} \Bigg)+c~.
 \ee
 The integration constant can be fixed by using the fact that in the vacuum $U=V=0$ and for the constant time interval $l_u=l_v$, the EE has no contribution from CS part. Thus $ c=0 $ and
  \be
 S_\text{CS} =\frac{1}{4G_N \mu}  \log  \Bigg( \frac{ \sqrt V \sinh  (l_u\sqrt U)}{\sqrt U  \sinh  (l_v\sqrt V)} \Bigg) ~.
 \ee
 which agrees with \eqref{EECSnew}. 
 
 Of course, the Einstein-Hilbert contribution can also be evaluated in this way. This method is quite general and the key point is to regard the entanglement entropy, like black hole entropy, as Noether charge. The Noether charge can be derived systematically using covariant phase space formalism and it applies to all the gravitational system. Once obtaining the  variation of Noether charge, we can integrate in phase space and get the entanglement entropy.  So this is a quite universal approach to calculate EE. It is interesting to apply this method to other holographic systems. 
 

\bibliographystyle{apsrev4-1} 
\bibliography{TMG}

\end{document}

%% file: ShortCommand.tex
\newcommand{\be}{\begin{equation}}
\newcommand{\ee}{\end{equation}}
\newcommand{\bpm}{\begin{pmatrix}}
\newcommand{\epm}{\end{pmatrix}}

\newcommand{\EV}[1]{\langle #1 \rangle}
\newcommand{\beqn}{\begin{eqnarray}}
\newcommand{\eeqn}{\end{eqnarray}}

\newcommand{\p}{\partial}

\DeclareMathOperator{\tr}{tr}